\documentclass[apjs]{emulateapj}
\usepackage{natbib}
\usepackage[colorlinks,citecolor=blue,urlcolor=blue]{hyperref}
\usepackage{breakurl}
\usepackage{amsmath}
\usepackage{amssymb}
\usepackage{graphicx}

\def\apjl{{ApJ}}                
\def\apjs{{ApJS}}               
\def\ao{{Appl.~Opt.}}           
\def\aap{{A\&A}}                

\newcommand{\sersic}{S\'ersic}
\newcommand{\Reff}{\ensuremath{R_{\mathrm{eff}}}}

\newcommand{\rev}[1]{#1}
\newcommand{\galsimpaper}{Rowe et~al., {\em in prep.}}

\shorttitle{GREAT3 Handbook}
\shortauthors{Mandelbaum, Rowe, et al.}
\begin{document}

\title{The Third Gravitational Lensing Accuracy Testing (GREAT3) Challenge Handbook}

\author{
Rachel Mandelbaum\altaffilmark{1},
Barnaby Rowe\altaffilmark{2,}\altaffilmark{3},
James Bosch\altaffilmark{4},
Chihway Chang\altaffilmark{5},
Frederic Courbin\altaffilmark{6},
Mandeep Gill\altaffilmark{5},
Mike Jarvis\altaffilmark{7},
Arun Kannawadi\altaffilmark{1},
Tomasz Kacprzak\altaffilmark{2},
Claire Lackner\altaffilmark{8},
Alexie Leauthaud\altaffilmark{8},
Hironao Miyatake\altaffilmark{4},
Reiko Nakajima\altaffilmark{9},
Jason Rhodes\altaffilmark{3,}\altaffilmark{10},
Melanie Simet\altaffilmark{1},
Joe Zuntz\altaffilmark{11},
Bob Armstrong\altaffilmark{7},
Sarah Bridle\altaffilmark{11},
Jean Coupon\altaffilmark{12},
J\"org P. Dietrich\altaffilmark{13,}\altaffilmark{14},
Marc Gentile\altaffilmark{6},
Catherine Heymans\altaffilmark{15},
Alden S. Jurling\altaffilmark{16,}\altaffilmark{17},
Stephen M. Kent\altaffilmark{18},
David Kirkby\altaffilmark{19},
Daniel Margala\altaffilmark{19},
Richard Massey\altaffilmark{20},
Peter Melchior\altaffilmark{21},
John Peterson\altaffilmark{22},
Aaron Roodman\altaffilmark{5},
\& Tim Schrabback\altaffilmark{9}
}

\altaffiltext{1}{McWilliams Center for Cosmology, Carnegie Mellon
  University, 5000 Forbes Avenue, Pittsburgh, PA 15213, USA} 
\altaffiltext{2}{Department of Physics and Astronomy, University
  College London, Gower Street, London WC1E 6BT, UK} 
\altaffiltext{3}{Jet Propulsion Laboratory, California Institute of
  Technology, MS 300‐315, 4800 Oak Grove Drive, Pasadena, CA} 
\altaffiltext{4}{Department of Astrophysical Sciences, Princeton
  University, Princeton, NJ 08544, USA} 
\altaffiltext{5}{KIPAC, Stanford University, 452 Lomita Mall,
  Stanford, CA 94309, USA} 
\altaffiltext{6}{Laboratoire d'astrophysique, Ecole Polytechnique
  F\'ed\'erale de Lausanne (EPFL), Observatoire de Sauverny, CH-1290
  Versoix, Switzerland} 
\altaffiltext{7}{Department of Physics and Astronomy, University of
  Pennsylvania, 209 South 33rd Street, Philadelphia, PA 19104, USA} 
\altaffiltext{8}{Kavli Institute for the Physics and Mathematics of
  the Universe (WPI), Todai Institutes for Advanced Study, the
  University of Tokyo, Kashiwa, Japan} 
\altaffiltext{9}{Argelander-Institut f\"ur Astronomie, Universitat
  Bonn, D-53121 Bonn, Germany} 
\altaffiltext{10}{California Institute of Technology, MC 350-17, 1200
  East California Boulevard, Pasadena, CA 91125, USA} 
\altaffiltext{11}{Jodrell Bank Centre for Astrophysics, School of
  Physics and Astronomy, The University of Manchester, Manchester M13
  9PL, UK} 
\altaffiltext{12}{Institute of Astronomy and Astrophysics, Academia Sinica, 
P.O. Box 23-141, Taipei 10617, Taiwan} 
\altaffiltext{13}{Universit\"ats-Sternwarte M\"unchen, Scheinerstr. 1, 81679 M\"unchen,
Germany} 
\altaffiltext{14}{Excellence Cluster Universe, 85748 Garching
  b. M\"unchen, Germany} 
\altaffiltext{15}{Scottish Universities Physics Alliance, Institute
  for Astronomy, University of Edinburgh, Royal Observatory, Blackford
  Hill, Edinburgh, EH9 3HJ, UK} 
\altaffiltext{16}{NASA Goddard Space Flight Center, Greenbelt, MD
  20771} 
\altaffiltext{17}{Department of Physics and Astronomy, University of
  Rochester, Rochester NY, 14618, USA} 
\altaffiltext{18}{Center for Particle Astrophysics, Fermi National
  Accelerator Laboratory, Batavia, IL 60510, USA; operated by Fermi Research Alliance,
    LLC under Contract No. De-AC02-07CH11359 with the United States
    Department of Energy.} 
\altaffiltext{19}{Department of Physics and Astronomy, University of
  California, Irvine, CA 92697, USA} 
\altaffiltext{20}{Institute for Computational Cosmology, Durham
  University, South Road, Durham DH1 3LE, UK} 
\altaffiltext{21}{Center for Cosmology and Astro-Particle Physics \&
  Department of Physics, The Ohio State University, Columbus, OH,
  43210, USA}  
\altaffiltext{22}{Department of Physics, Purdue University, West
  Lafayette, IN 47907, USA} 

\begin{abstract}
  The GRavitational lEnsing Accuracy Testing 3 (GREAT3) challenge is
  the third in a series of image analysis challenges, with a goal of
  testing and facilitating the development of methods for analyzing
  astronomical images that will be used to measure weak gravitational
  lensing.  This measurement requires extremely precise estimation of
  very small galaxy shape distortions, in the presence of far larger
  intrinsic galaxy shapes and distortions due to the blurring kernel
  caused by the atmosphere, telescope optics, and instrumental
  effects.  The GREAT3 challenge is posed to the astronomy, machine
  learning, and statistics communities, and includes tests of three 
  specific effects that are of immediate relevance to upcoming weak
  lensing surveys, two of which have never been tested in a community
  challenge before.  These effects include \rev{many novel aspects
    including} realistically
  complex galaxy models based on high-resolution imaging from space;
  spatially varying\rev{, physically-motivated} blurring kernel; and combination of 
  multiple different
  exposures.  To facilitate entry by people new to the field, and for
  use as a diagnostic tool, the
  simulation software for the challenge is publicly available, though
  the exact parameters used for the challenge are
  blinded.  Sample scripts to analyze the challenge data using
  existing methods will also be provided.  See 
  \url{http://great3challenge.info} and \url{http://great3.projects.phys.ucl.ac.uk/leaderboard/} for more information.
\end{abstract}


\section{Introduction}\label{sec:intro}

In our currently accepted cosmological model, the baryonic matter from
which stars and planets are made accounts for only 4\% of the energy
density of the Universe.  In order to explain many 
cosmological observations, we have been forced to posit the existence
of dark matter (which we detect through its gravitational attraction)
and dark energy (which causes a repulsion that is driving the 
accelerated expansion of the Universe, the discovery of which led to
the 2011 Nobel Prize in Physics).  While we infer the existence of
these dark components, the question of what they actually are remains
a mystery.

Gravitational lensing is the deflection of light from distant objects
by all matter along its path, including dark
matter (Fig.~\ref{fig:lensing-intro}).  Lensing
measurements are thus directly sensitive to dark matter.  They also
permit us to infer the properties of dark energy \citep{2002PhRvD..66h3515H},
because the accelerated expansion of the Universe that it causes
directly opposes the effects of gravity (which tends to cause matter
to clump into ever larger structures) and influences light propagation through its impact on the geometry of the universe.

This measurement entails detecting small but spatially coherent
distortions (known as weak shears) in the shapes of distant galaxies,
which provide a statistical map of large-scale cosmological
structures. Weak lensing measurements have already placed some 
constraint on the growth of structure, typically with 10\%
statistical errors, or as small as 5\% for the most recent analyses
\citep{2013MNRAS.432.2433H,2013ApJ...765...74J}. Because of the
sensitivity of weak lensing to the dark components of
the Universe, the astronomical community has designed 
upcoming surveys to measure it very precisely, and thereby
constrain cosmological parameters. In addition to several experiments
beginning in 2013, there are even larger experiments
that are planned to start at the end of this decade. In the Astro2010 Decadal Survey of
US astronomy \citep{decadal}, the most highly endorsed large experiments both from the
ground 
(the Large Synoptic Survey Telescope, or LSST) and space (the
Wide-Field Infrared Survey Telescope, or WFIRST-AFTA) are ones with a
significant emphasis on weak lensing cosmology.  The European Space
Agency recently decided that of several possible large space-based
astronomical surveys, they will proceed with the Euclid mission, which
likewise has a major emphasis on lensing.

However, the increasing size of these experiments, and the decreasing
statistical errors, comes with a price: to fully realize 
their promise, we must understand systematic errors increasingly
well.  The coherent lensing distortions of galaxy shapes
are typically $\sim 1$\% in size, far smaller than galaxy
intrinsic ellipticities ($\sim 0.3$) and, more problematically,
smaller than the coherent distortions due to light propagation
through the atmosphere and telescope optics (the
point-spread function, or PSF).  Removing the effects of the PSF and
measuring lensing shears for galaxies that are only
moderately resolved and have limited signal-to-noise is a demanding
statistical problem that has not been solved adequately for
upcoming surveys.  Systematic errors related to shape measurement must
be reduced by factors of 5-10 in the next decade.  The weak lensing
community has gained substantially from a practice of carrying out blind challenges in order to
test shear measurement methods.  By using
simulated data in which the ground truth is known, but with
realistically complicated galaxies and PSFs, we can 
estimate the systematic errors associated with current methods, and
use our new knowledge of their failure modes to spur further
development in the field.

\subsection{Previous challenges and the context for GREAT3}

The current and past GREAT challenges have all been supported by the
PASCAL network.  
The GREAT08 Challenge
\citep{2009AnApS...3....6B,2010MNRAS.405.2044B} set a highly
simplified version of the problem, using known PSFs, simple galaxy
models, and a constant applied gravitational shear.  The GREAT10
Challenge
\citep{2010arXiv1009.0779K,2012MNRAS.423.3163K,2013ApJS..205...12K}
increased the realism and complexity of its simulations over GREAT08
by using cosmologically-varying shear fields and greater
variation in galaxy model parameters and telescope observing
conditions. Since imperfect knowledge of the PSF can also bias shear
measurements, GREAT10 tested PSF modelling in a standalone ‘Star
Challenge’. GREAT08 and GREAT10 were preceded by a number of
internal challenges within the astrophysics community, known as
the Shear Testing Programme, or STEP
\citep{2006MNRAS.368.1323H,2007MNRAS.376...13M}, which demonstrated
the highly non-trivial nature of the shear measurement problem.  Both
GREAT08 
and GREAT10 generated significant (factors of 2-3) improvement in the
accuracy of weak lensing shape measurement, while also 
providing a greater understanding of the major limitations of
existing methods.

The key goals of the GREAT3 challenge are to facilitate further work in
understanding existing methods of PSF correction, to suggest ways that
they can be developed and improved in the future, and to spur the
creation of new methods that solve the limitations of existing ones.
We aim to address the
challenges in this field in two ways:
(1) We provide a suite of simulated galaxy images for making controlled tests
of outstanding issues in lensing shear 
measurement, focusing on crucial issues not addressed in previous
challenges and adding new levels of realism; and (2) we provide the
simulation code, GalSim (\galsimpaper), as a
fully documented, open source (licensed under the GNU General Public License,
or GPL) development toolkit in a modern language
framework (object-oriented Python wrapping around C++, \S\ref{sec:sims}).
GalSim is already public\footnote{\url{https://github.com/GalSim-developers/GalSim}}, but the exact challenge input
parameters are blinded.  We anticipate that the open source status of this
simulation code will facilitate more rapid improvement of existing
methods and development of new ones.  Using real galaxy images
(from the {\em Hubble Space Telescope}) and detailed, physically
motivated PSF models as inputs will ensure that the space of possible
simulations is overwhelmingly large, as in reality.  This development will be
crucial for weak lensing to achieve its unique potential for
understanding the nature of dark energy and matter in the Universe. In
this way we 
accurately reflect the real problem of galaxy shape measurement,
which combines well understood gravitational physics with 
input galaxies and observing conditions about which we have significantly less information.   

As for previous lensing challenges, the simulations are 
statistically matched to the size of the largest upcoming weak lensing
surveys such as Euclid\footnote{\url{http://sci.esa.int/euclid}}
\citep{2011arXiv1110.3193L},
LSST\footnote{\url{http://www.lsst.org/lsst/}}
\citep{2009arXiv0912.0201L}, and
WFIRST-AFTA\footnote{\url{http://wfirst.gsfc.nasa.gov}}
\citep{2013arXiv1305.5422S}.  The goal of participants is to measure
gravitational shears sufficiently precisely that systematic errors in
the measurements are below the statistical errors,
so that the error budget is not dominated by 
systematics.  The challenge is split into branches that
reflect different issues in the field and types of observations;
participants may enter as many or as few branches as they
wish. \rev{The simulation design includes many new aspects relative to
previous challenges, as is required in order for us to carry out tests
of several important issues in the field outlined in \S\ref{sec:issues}.}

An overview of how to use this handbook for the GREAT3 Challenge
is as follows.
\S\ref{sec:physics} presents an overview of the physics behind gravitational lensing and
astronomical imaging, to motivate the major issues in shear estimation.
To estimate the gravitational shear in the galaxy image,
the standard procedure is to measure each galaxy shape and infer the
overall shear from these; an overview of existing approaches to
shape measurement is in Appendix~\ref{sec:shearmeas}.
\S\ref{sec:issues} contains a summary of the specific issues that the GREAT3
Challenge is designed to address.  The structure of the
Challenge and how it is run is detailed in \S\ref{sec:challenge}. In \S\ref{sec:sims} we describe
the simulation generation and design, and relate it to 
the issues from \S\ref{sec:issues}.  
Finally, we summarize
the simplifications of the GREAT3 challenge in \S\ref{sec:future}.

\section{Physics background}\label{sec:physics}

\begin{figure*}
\begin{center}
\includegraphics[width=1.8\columnwidth,angle=0]{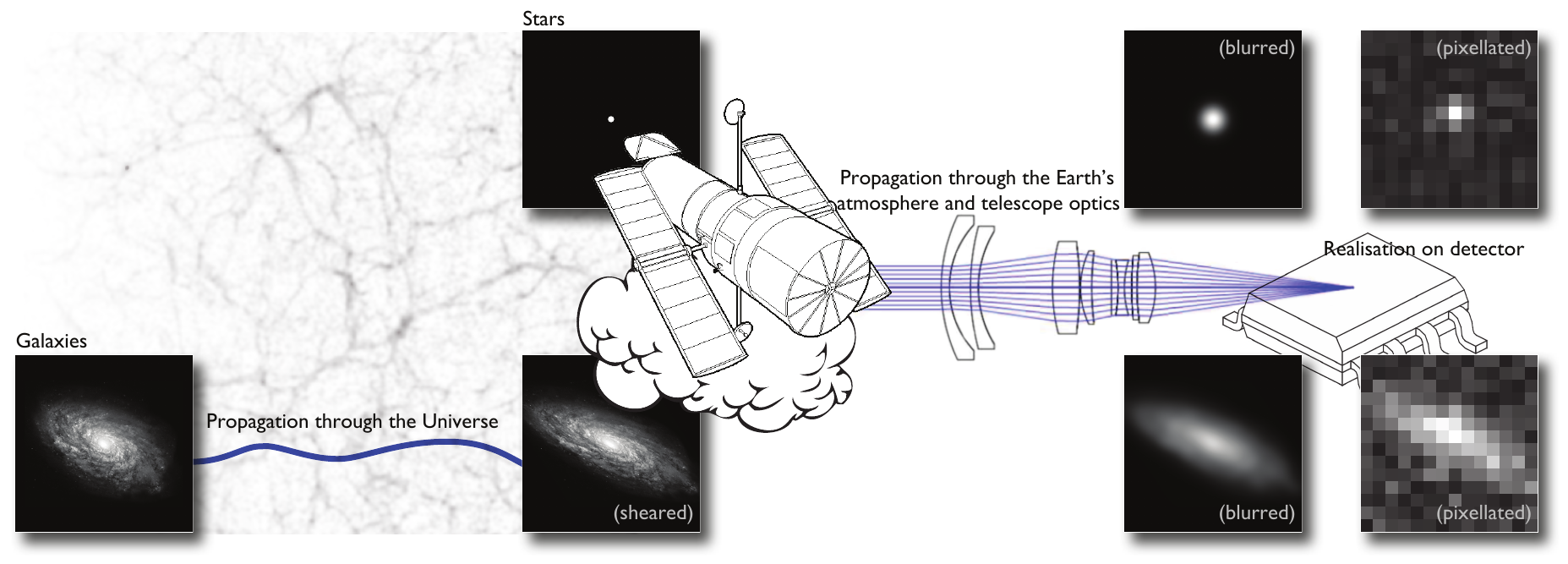}
\caption{\label{fig:lensing-intro}An illustration of the process of
  gravitational lensing and other effects that change the apparent
  shapes of galaxies in the astronomical imaging process.  (Based on Figure~8
  from \citealt{2010arXiv1009.0779K}).}
\end{center}
\end{figure*}

Here we describe the basic physics behind gravitational lensing
and astronomical imaging.  The processes described in this
section are 
shown in Fig.~\ref{fig:lensing-intro}.

\subsection{Lensing shear}\label{subsec:physics:shear}

Gravitational lensing distorts observed images of distant galaxies, in
a way that depends on the distribution of mass around the line of
sight.  This distortion can be described as a general coordinate
transformation, but for the overwhelming majority of distant galaxy light
sources, the transformation is well approximated as being locally
linear. This limit is known as weak gravitational lensing. 

Weak gravitational lensing can be described as a linear
transformation between unlensed coordinates ($x_u$, $y_u$; with the
origin at the center of the distant light source) and the
lensed coordinates in which we observe galaxies ($x_l$, $y_l$; with the origin at the center of the
observed image),
\begin{equation}\label{eq:lensingshear}
  \left( \begin{array}{c} x_u \\ y_u \end{array} \right)
  = \left( \begin{array}{cc} 1 - \gamma_1 - \kappa & - \gamma_2 \\ -\gamma_2 & 1+\gamma_1 - \kappa \end{array} \right)
  \left( \begin{array}{c} x_l \\ y_l \end{array} \right).
\end{equation}
Here we have introduced the two components of the complex-valued lensing shear
$\gamma = \gamma_1 + {\rm i}\gamma_2$, and the lensing convergence $\kappa$.  The
shear describes the \emph{stretching} of galaxy images due to
lensing.  The convention is such that a positive (negative) $\gamma_1$ results in an image being
stretched along the $x$ ($y$) axis direction.  Likewise a positive (negative) $\gamma_2$
results in an image being stretched along the line $y=x$ ($y=-x$).  The convergence
$\kappa$ describes a change in apparent size for lensed objects: areas
of the sky for which $\kappa$ is positive have apparent changes in
area (at fixed surface brightness) that make lensed images appear
larger and brighter than if they were unlensed, and a modified galaxy
density.

Often, as we do not know the distribution of sizes of distant galaxies
well, it is common to recast the transformation
\eqref{eq:lensingshear} as
\begin{equation}
\left( \begin{array}{c} x_u \\ y_u \end{array} \right)
= (1 - \kappa) \left( \begin{array}{cc} 1 - g_1 & - g_2 \\ -g_2 & 1+g_1\end{array} \right)
\left( \begin{array}{c} x_l \\ y_l \end{array} \right),
\end{equation}
in terms of the \emph{reduced} shear,
$g_1 = \gamma_1 / (1 - \kappa)$ and $g_2 = \gamma_2 / (1 - \kappa)$.
In many applications the $(1 - \kappa)$ term is not estimated from the
data (although see, e.g., \citealp{2013MNRAS.430.2844C}), and so it is the image
stretching described by the reduced shear that is in fact observed in
galaxies (hence the use of this notation in \citealp{2009AnApS...3....6B}).  
We often encode the two components of shear and reduced shear into a
single complex number, e.g.\ $\gamma = \gamma_1 + {\rm i}\gamma_2$, $g = g_1
+ {\rm i} g_2$.  In most cosmological applications $g \simeq \gamma$ is a
reasonable approximation; however, the GREAT3
simulations with cosmologically varying shear fields do also contain a
corresponding $\kappa$ variation.

The lensing shear causes
a change in estimates of the \emph{ellipticity} of distant galaxies.
If sources with intrinsically circular isophotes (contours of equal brightness) could be
identified, the observed sources (post-lensing) would have elliptical isophotes that
we can characterize by their minor-to-major axis ratio $b / a$
and the orientation of the major axis $\phi$.
For $|g|<1$, these directly yield a value of the reduced shear
\begin{equation}\label{eq:gab}
|g| = \frac{1 - b/a}{1 + b / a}
\end{equation}
which, combined with the orientation $\phi$, gives the two orthogonal
components of shear $g_1 = |g| \cos{2 \phi}$, $g_2 =
|g| \sin{2 \phi}$.

In practice we cannot identify distant galaxy sources that are
circular prior to lensing, nor do distant galaxies have elliptical
isophotes.  However, it is possible to estimate properties that
transform in similar ways to the simplified case presented above, and
from which we can extract statistical estimates of shear.  One method
is to model the light from galaxies using a profile that does have a
well-defined ellipticity.  We can write this ellipticity as a complex
number $\varepsilon = \varepsilon_1 + {\rm i} \varepsilon_2$, with
magnitude $|\varepsilon| = (1 - b/a) / (1 + b/a)$ and orientation angle
determined by the direction of the major elliptical axis.  Under an applied shear with $|g| \le
1$, this definition of ellipticity transforms as
\begin{equation}\label{eq:gsheartrans}
\varepsilon = \frac{\varepsilon^{(s)} + g}{1 + g^* \varepsilon^{(s)} }
\end{equation}
(see \citealp{2001PhR...340..291B} for the strong shear $|g| > 1$ result).
Here we have labelled the ellipticity of the source prior to lensing
as $\varepsilon^{(s)}$.  For $g\ll 1$, eq.~\eqref{eq:gsheartrans} becomes $\varepsilon \simeq
\varepsilon^{(s)} + g$.  For a population of source ellipticities that
are randomly oriented so that $ \langle \varepsilon^{(s)} \rangle =
0$, the ensemble average ellipticity after lensing
gives an unbiased estimate of the shear: $\langle \varepsilon \rangle
\simeq g $.

Another common choice of shape parametrization is based on second brightness moments across the
galaxy image,
\begin{equation}\label{eq:qij}
Q_{ij} = \frac{\int {\rm d}^2 x I({\bf x}) W({\bf x}) x_i x_j }
{\int {\rm d}^2 x I({\bf x}) W({\bf x}) },
\end{equation}
where the coordinates $x_1$ and $x_2$ correspond to the $x$ and $y$
directions respectively, $I({\bf x})$ denotes the galaxy image light
profile, $W({\bf x})$ is an optional\footnote{Optional for the purpose
  of this definition; but in practice, for images with noise, some
  weight function that reduces the contribution from the wings of the
  galaxy is necessary to avoid moments being dominated by noise.}
weighting function (see \citealp{schneider06}), and where  
the coordinate origin ${\bf x} = 0$ is placed at the galaxy
image center (commonly called the centroid). A second definition
of ellipticity, sometimes referred to as the \emph{distortion} to
distinguish it from the ellipticity that satisfies equation
\eqref{eq:gsheartrans}, can then be written as
\begin{equation}\label{eq:ellipticity}
e = e_1 + {\rm i} e_2 = \frac{Q_{11} - Q_{22} + 2 {\rm i} Q_{12}}{Q_{11} + Q_{22}}.
\end{equation}
The ellipticity $\varepsilon$ can also be related to the moments, like
the distortion, but replacing the denominator in
Eq.~\ref{eq:ellipticity} with $Q_{11} + Q_{22} + 2 (Q_{11}Q_{22}-Q_{12}^2)^{1/2}$.

If the weighting function $W=1$ (unweighted moments) or $W=W[I({\bf
  x})]$ (a brightness-dependent weight function) then an image with
elliptical isophotes of axis ratio $b/a$ has
\begin{equation}
|e| = \frac{1 - b^2 / a^2}{1 + b^2 / a^2}.
\end{equation}
Under a shear, $e$ transforms from a source (pre-lensing)
distortion $e^{(s)}$ as 
\begin{equation}
e = \frac{e^{(s)} + 2g + g^2e^{(s) *}}{1 + |g|^2 + 2 \Re[g e^{(s)*}]},
\end{equation}
so that in the weak shear limit, $e \simeq e^{(s)} + 2 [1-(e^{(s)})^2]g$.
For a population of source distortions that
are randomly oriented so that $ \langle e^{(s)} \rangle =
0$, the ensemble average $e$ after lensing
gives an unbiased estimate of approximately twice the shear that depends on the
population root mean square (RMS) ellipticity, $\langle e \rangle
\simeq 2[1-\langle (e^{(s)})^2\rangle] g $.

\subsection{Shear fields}\label{subsec:shearfields}

Although gravitational lensing distortions at the locations of
individual galaxies can typically be approximated as linear, the shear
and convergence vary with position across the sky.  This
variation is due to the non-uniform distribution of massive structures
in the universe.   Estimates of this variation, which are discrete
estimates of the underlying shear field, are used in various ways to
improve our models of the universe.  In the following Section we will
primarily focus on shear fields\footnote{See e.g.\ 
\citet{2011arXiv1111.1070H,2012ApJ...744L..22S}, \&
\citet{2013MNRAS.430.2844C} (and references therein) 
for recent developments in 
lensing magnification.}.

One well-motivated place to look for shears is around structures 
that can be directly observed, for example around galaxies likely
to lie in dark matter-rich regions (a study known as \emph{galaxy-galaxy
  lensing}), or around foreground galaxy clusters.
Around a central lens object, the tangential shear induced by gravitational lensing
is approximately constant at fixed radius; thus measuring constant
shear is a goal of galaxy-galaxy and cluster-galaxy lensing measurements.  Estimates of the
shear around such objects have been compared to parametric models of
the matter content to provide great insight into the way that
visible matter is traced by underlying mass.

\begin{figure*}
\begin{center}
\includegraphics[width=1.3\columnwidth,angle=0]{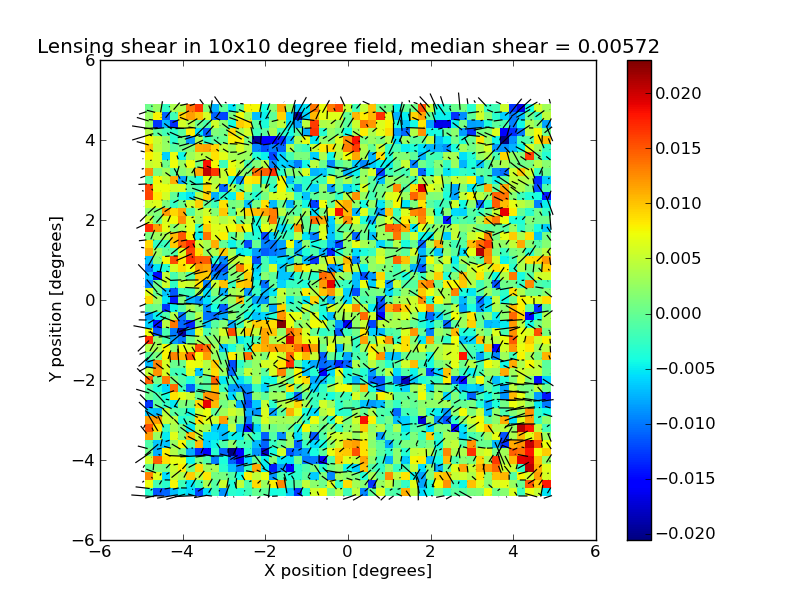}
\caption{\label{fig:shear_field} An example of a cosmological shear
  field in a  $10\times 10$
deg$^2$ region of sky, with the same statistical properties as
realistic cosmological shear fields.  At each point on the grid, the
size of the arrow shows the magnitude of the shear (for reference, the plot title
gives the median shear value), the orientation shows
the shear direction at that location, and the color shows the
convergence $\kappa$.  As shown, the shear field 
exhibits coherent alignments over large scales, with tangential
orientation around mass overdensities.} 
\end{center}
\end{figure*}

The statistics of lensing shear, and its spatial correlation as a
function of angular scale on the sky, is another key prediction of
many models of the universe, and this is therefore the other goal of our
measurement.  Comparison of these models to 
observations is expected to greatly improve our understanding of
dark matter and dark energy, and to constrain models of modified
gravity that have been posed as another way to explain the accelerated
expansion of the Universe (e.g., \citealp{2006ewg3.rept.....P,2009arXiv0901.0721A}).
Fig.~\ref{fig:shear_field} shows a typical cosmological shear field in a $10\times 10$
deg$^2$ region (the size of GREAT3 images).

A map of galaxy shears is a spin-2 (headless vector) field.  As such, 
it can be decomposed into two components, called the E-mode and B-mode
by analogy with electric and magnetic fields.  The E-mode can be represented
as the gradient of a scalar field, and has no curl; the B-mode can be
represented as the curl of a vector field, and has no divergence.

Cosmological shear fields include
almost exclusively $E$ mode signals - with tangential shears induced
around density peaks - and only negligible $B$ modes\footnote{Some B-mode contributions can occur due to 
multiple lens planes, source clustering, and other effects, all of which are
quite small \citep[e.g.,][]{2002A&A...389..729S,2004PhRvD..70f3526H,2004ApJ...613L...1V}.}, a
fact that is often exploited in reality to test for systematic errors
(observed $B$ modes are taken as a sign of systematic error).  

\subsection{Measuring shear fields}

The shear fields around galaxies and galaxy clusters  are
generally used to constrain either parametric models of the unseen
mass distribution around these objects  (e.g.,
\citealp{2012ApJ...744..159L,2013arXiv1304.4265V}),
 or a non-parametric map of the same (e.g.,
\citealp{2012Natur.487..202D,2013MNRAS.433.3373V}).  When estimating shears
around astrophysical objects for the purpose of fitting a parametric
model, it is common to estimate the average shear in annuli of 
separation from the  center of the foreground object.  For
more general map making, shear estimates from individual
galaxies are typically averaged in cells across the sky, and the
smoothed shears are then used to estimate the projected density.

However, another important application of weak lensing shear estimates
is to probe the statistical properties of the shear field, as a
function of angular separation on the sky.  Different models of the
universe predict differing statistical distributions of shear as a
function of angular scale.  Recent estimates of the spatial
correlations between shears, and the evolution of these correlations over
cosmic time
\citep{2013ApJ...765...74J,2013MNRAS.432.2433H}, used catalogs
of shear estimates across a whole survey area as a probe of the
growth of matter structure.
In these `cosmic shear' analyses, the shear correlation function
estimated from catalogs of shears
(see Appendix \ref{sec:cf}) is the data vector for
the estimation of cosmological parameters.

The shear power spectrum, related to the correlation function by
a Fourier transform, is also a quantity of interest for describing the
statistics of cosmic shear.
In the GREAT10 challenge, the goal was to estimate the
power spectrum directly from the shear using the discrete Fourier
transform due to the galaxies being positioned at fixed grid
locations.  However, in practice the use of power 
spectra presents challenges because of the non-regular spacing
of galaxies on the sky, and the presence of holes in coverage
due to bright foreground objects or camera defects.  In GREAT3, we
adopt a correlation-function based metric for the 
simulations containing variable shear.  As a useful side product, this
also allows us to sensitively probe contamination of the shear field
due to both variable and constant point-spread functions (PSFs).  In
the following Section we discuss the importance of the PSF in weak
lensing science.

\subsection{Variable PSF}\label{subsec:varpsf}

After the light from distant galaxies is 
sheared, it passes through the turbulent atmosphere (when observing with a
ground-based telescope), and through telescope optics and a detector.
While the initial shear is the desired signal,
these later effects (which can typically be modeled as convolution with a
blurring kernel called the PSF) systematically modify the images.  The
blurring due to the atmosphere is typically 
larger than that due to optics, and varies relatively rapidly in time
compared to typical exposure times for astronomical imaging
\citep[e.g.,][]{2012MNRAS.421..381H, 2013arXiv1304.4992H}.  In
contrast, the PSF due to the optics varies relatively slowly with time.  The optical PSF
is commonly described as a combination of diffraction 
plus aberrations (possibly up to quite high order).   Both the
atmospheric and optical PSF have some spatial coherence, qualitatively
like lensing shear, though the scaling with separation is not identical.

The effect of the PSF on the galaxy shapes that we wish to measure is
twofold: first, applying a roughly circular blurring kernel tends to
dilute the galaxy shapes, making them appear rounder by an amount
that depends on the ratio of galaxy and PSF sizes. 
Correction for this  dilution can easily be a factor of 2 for
typical galaxies, for which we wish to measure shears to 1\%.  Second,
the small but coherent PSF anisotropies can leak into the galaxy shapes if not
removed, mimicking a lensing signal.

Stars in the images
are effectively point sources before blurring by the PSF, and hence
are measures of the PSF.  However, the PSF must be estimated from them
and then interpolated to the positions of galaxies.  For a summary of
some common methods of PSF estimation and interpolation, see
\cite{2013ApJS..205...12K}.

\subsection{Summary of effects}

Fig.~\ref{fig:lensing-intro} summarizes the main effects that go into
a weak lensing observation.  The galaxy image is distorted as it is
deflected by mass along the line-of-sight from the galaxy to us.  This is the desired signal.
It is then further distorted by the atmosphere
(for a ground-based telescope), telescope optics, and pixelation on
the detector; these effects collectively form the PSF and are
equivalent to convolution\footnote{This equivalence is valid in the
  limit that these functions are continuous.  For data that are
  discretely sampled, it is important to ensure that they are Nyquist
  sampled, otherwise the statement that pixelation can be treated as a
  convolution is false.} with a blurring kernel.  The images have 
noise, which can cause a bias when solving the non-linear problem of
estimating the original shear, and there are also detector 
effects (not shown here).  Given that upcoming datasets will have
hundreds of million or billions of galaxies, 
removing these nuisance effects to sub-percent precision
is a necessary but formidable challenge.

\section{Important  issues in the field}\label{sec:issues}

The goal of this challenge is to address three major open issues in
the field of weak lensing, as determined by a consensus among the
community.  These could conceivably be
limiting systematic errors for weak lensing surveys beginning this year, but their
importance has not been systematically quantified in a community
challenge.  In the interest of making a fair test of these issues, we
exclude other issues that were deemed to be of lesser importance for
now (\S\ref{sec:future}).  The GREAT3 challenge consists of
experiments 
that can test each of the issues below separately, so that people who
are interested in only certain issues can still participate.

\subsection{Realistic galaxy morphologies}\label{subsec:issues:realgal}

Multiple studies have shown that no method of shape measurement 
based on second moments can be completely independent of the
details of the galaxy population (e.g., morphology and substructure),
because the shear couples the second moments to the higher-order
moments
\citep{2007MNRAS.380..229M,2010MNRAS.406.2793B,2011MNRAS.414.1047Z}.
This issue is particularly pressing given that several 
state-of-the-art shape measurement methods (see 
Appendix~\ref{sec:shearmeas}) are based on fitting relatively simple galaxy
models or are based on a decomposition into basis functions that cannot
necessarily describe galaxy profiles in detail \citep{2010MNRAS.404..458V,
2010A&A...510A..75M}.  More complex decompositions into basis
functions often can describe more complex galaxies, but at
the expense of introducing many tens or $>100$ parameters, making them
impractical for typical images with typical signal-to-noise ratios
$S/N\sim 10$--$20$ (see \S\ref{subsec:sims:noise} for the formal
definition of this quantity). In addition,
methods that measure galaxy distortions
(\S\ref{subsec:physics:shear}) require an estimate of the intrinsic
RMS galaxy distortion to convert to an ensemble shear,
resulting in another type of dependency on the underlying nature of the galaxy
population.

\begin{figure}
\begin{center}
\includegraphics[width=0.9\columnwidth,angle=0]{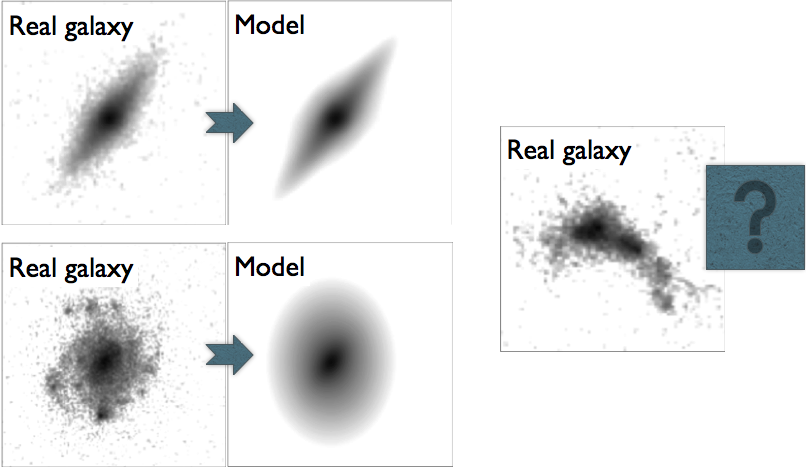}
\caption{\label{fig:real_gal}Real galaxies from the {\em HST} as
  observed by the Advanced Camera for Surveys (ACS) in the COSMOS
  survey
  \citep{2007ApJS..172..196K,2007ApJS..172....1S,2007ApJS..172...38S}.
  The top left shows a galaxy that is well-fit by a simple parametric
  model from \cite{2012MNRAS.421.2277L}.  The bottom left shows a
  galaxy that is reasonably well-fit but with additional substructure
  evident.  The right side shows a true ``irregular'' galaxy that
  is not well-fit by simple parametric models with  $\sim 10$
  parameters.}
\end{center}
\end{figure}

As an illustration of this problem, Fig.~\ref{fig:real_gal} shows
several typical galaxies in high-resolution data from the
{\em Hubble Space Telescope} ({\em HST}).  Only a few tens of percent ($\sim 20$\%) of
galaxies can be perfectly fit by simple galaxy models such as those 
commonly used by model-fitting methods today (e.g., top left); nearly half can be fit by them, but with 
additional substructure clearly evident (e.g., bottom left); and a few
tens of percent ($\sim 30$\%) are true ``irregulars'' that cannot 
be fit by simple models at all (right panel).
The GREAT08 and GREAT10 challenges  used simple galaxy models, which
motivates us to explore the impact of
realistic galaxy morphology on shape measurement in the GREAT3
challenge, thus constraining ``underfitting
biases''\footnote{These are biases in an $M$-parameter fit 
  that arise when the true image has $N>M$ parameters, and some of the
  $N-M$ additional parameters correlate with the shear; e.g., \citealt{2010MNRAS.406.2793B}.}. Nearly all
lensing data is lower resolution than what is shown in
Fig~\ref{fig:real_gal}; however, for this particular scientific
application, we have reasons to believe that what we do
{\em not}
know (the unresolved, detailed galaxy morphology) {\em does} hurt us at some unknown level.
One goal of the GREAT3 challenge is to quantify the extent to which
that statement is true.

The galaxies used for these simulations therefore come from {\em HST}. The technique for rendering the appearance of
these galaxies with an added lensing shear is
in \cite{2012MNRAS.420.1518M} and 
\S\ref{subsec:sims:galaxies} of this handbook.

\subsection{Variable PSFs}\label{subsec:issues:realpsf}

As discussed in \S\ref{subsec:varpsf}, realistic PSFs have complex profiles 
and spatial variation due to the turbulent atmosphere (in ground-based 
measurements) and the instrument (the telescope and the camera). 
Different approaches have been used to study these PSF characteristics 
using data and simulations \citep[e.g.,][]{2008arXiv0810.0027J, 
2012MNRAS.421..381H, 2013MNRAS.428.2695C, 2013arXiv1304.4992H}. 
We would like to test the impact of realistic PSFs on weak lensing
measurement, both for the case of (a) a realistically complex PSF
profile that is provided for participants, and (b) the case where the
PSF has spatial variation that the participants must infer from a
provided star field.  The latter test is complicated by the low
density of high-$S/N$ stars that can be used to infer the PSF, making
it hard to track high-frequency modes.

\begin{figure}
\begin{center}
\includegraphics[width=0.9\columnwidth,angle=0]{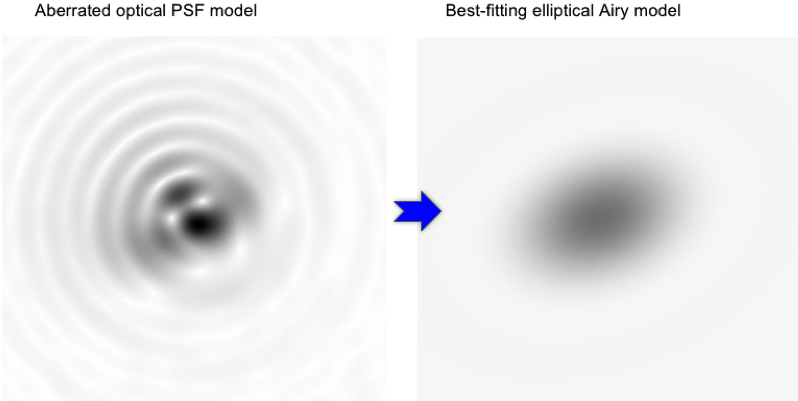}
\caption{\label{fig:real_psf}Examples showing the simplicity of
  common approximations to optical PSFs, i.e., without an atmospheric contribution.  \textbf{Left panel:}
  A realistic PSF, generated using lowest-order aberration
  theory with values that are typical for a well-aligned
  ground-based telescope, or for a (perhaps temporarily) misaligned
  space-based instrument.  \textbf{Right panel:} The best-fitting
  (least-squares) approximation to this PSF using an elliptical Airy
  disk, a parametrized PSF model used in the GREAT10 galaxy and
  star challenges.  The images are normalized to the same linear
  scale.}
\end{center}
\end{figure}

Case (a) can be motivated by Fig.~\ref{fig:real_psf}, which shows a
realistically complex PSF due to telescope optics and a simple
model that is commonly used to represent it.  As shown,
the former is more complicated than the latter, and it is plausible 
that shape measurement methods could behave differently for
the two cases.  For simulated data from ground-based telescopes, there
is also a convolution by the far broader atmospheric PSF\footnote{In
  the long-exposure limit, the PSF can simply be treated as the
  convolution of the optical and atmospheric PSF contributions.}.

Case (b), where the participants are required to infer the PSF, is
similar to the GREAT10 Star Challenge.  However, in that case the
participants were judged on the accuracy of their PSF reconstruction.
In GREAT3, the metric is the accuracy of shear field reconstruction,
i.e., we test how PSF determination errors propagate into the
recovered shear
field.  The value of this test is that different PSF
reconstructions at a particular RMS accuracy could actually involve
different spatial patterns in the residuals that affect shear field
recovery in different ways, so ultimately we must quantify the
performance of PSF estimation in terms of its impact on shear
measurement.

\subsection{Combination of multiple exposures}\label{subsec:issues:multexp}

Most datasets used for weak lensing measurement are not single
images, but rather multiple short exposures that are 
slightly offset from each other (``multi-epoch'' data).  Part of the data reduction procedure
involves combining them to estimate the galaxy shapes - either via
``co-addition'' to form a stacked image \citep[e.g. ][]{2012ApJ...761...15L, 
2013ApJ...765...74J}, or by applying some simultaneous fit procedure that 
treats each exposure separately \citep[e.g. ][]{2013MNRAS.429.3627M,
2013MNRAS.429.2858M}. Previous challenges have included  a single deep image.  In 
GREAT3, we include a test of how methods handle multiple images.

If the individual exposures are all Nyquist sampled and
there are no major distortions or holes in the data (due to defects,
cosmic rays, etc.), the combination of multi-exposure data is in
principle straightforward, making this test less interesting.
However, for a fraction of the data from ground-based telescopes, and
all data from upcoming space missions, the data in individual
exposures is {\em not} Nyquist sampled, which means that it is only
possible to create a critically-sampled image by combining the
multiple dithered (offset by sub-pixel amounts) images 
\citep[e.g.,][]{2011ApJ...741...46R}.  This is a more complicated algorithmic issue, 
and while our challenge does not address all aspects of it (e.g., it is even 
more complicated when there are holes in the data) we make a basic 
test of image combination.

When the PSF is very different in
some exposures than others, it is
possible to imagine gaining an
advantage by up-weighting higher-resolution data.  Hence it is
possible that a method that does the most basic, fundamentally
correct image combination could do worse than a method that is more clever
in how the exposures are combined.

\subsection{Challenge philosophy}\label{subsec:philosophy}

The GREAT3 challenge is structured as a series of 
experiments to evaluate three key issues 
separately before combining them.  Since our goal is
to address how important these issues are for extant
shape measurement methods (and encourage the development of new
methods that might address these issues better), we deliberately omit
some complications that were not chosen by the GREAT3 collaboration as
top priorities. For a list of omitted issues, see
\S\ref{sec:future}, and note that the simulation software is capable of
generating simulations that can address most of them.

One important note is the increased complexity compared to
GREAT08 and GREAT10, for which the simpler questions being asked
demanded simulations with (typically) $\delta$-function distributions
in galaxy and/or PSF 
parameters.  If a GREAT3 participant needs 
simulations at that level of simplicity to test their code, they can either download the GREAT08 or GREAT10
simulations, or generate new (simple) simulations with
 public simulation software.  Thus we are deliberately
including more realistic distributions of galaxy parameters, but still
in a format that allows for controlled experiments of the
impact of realistic galaxy morphology, real PSFs and their
variation, and combination of multiple exposure data.  The challenge 
structure described below (\S\ref{subsec:sim:branches}) reflects
this goal.

Like previous GREAT challenges, GREAT3 is meant 
to be inclusive of different data types.  In that spirit, it 
includes both ground- and space-based data (of which participants may
choose to analyze either or both); within those datasets, the images
have some variations of key parameters so that they do not 
appear to come from the same instrument.  Likewise, it has both
constant- and variable-shear data, as some methods have
assumptions that favor one or the other, and both are scientifically
useful.

\section{The challenge}\label{sec:challenge}

\subsection{Branch structure}\label{subsec:sim:branches}
To achieve the goals outlined in \S\ref{sec:issues}, the GREAT3 challenge
consists of five experiments:
\begin{itemize}
\item The control, which includes none of the three effects in
  \S\ref{sec:issues} (but is
  a non-trivial test of how shear estimation methods cope with a
  galaxy sample that has a continuous distribution in size and
  $S/N$).
\item Three experiments that each include only one of the effects of
  interest.
\item One experiment that includes all three effects together. 
\end{itemize}

\begin{figure*}
\begin{center}
\includegraphics[width=1.5\columnwidth,angle=0]{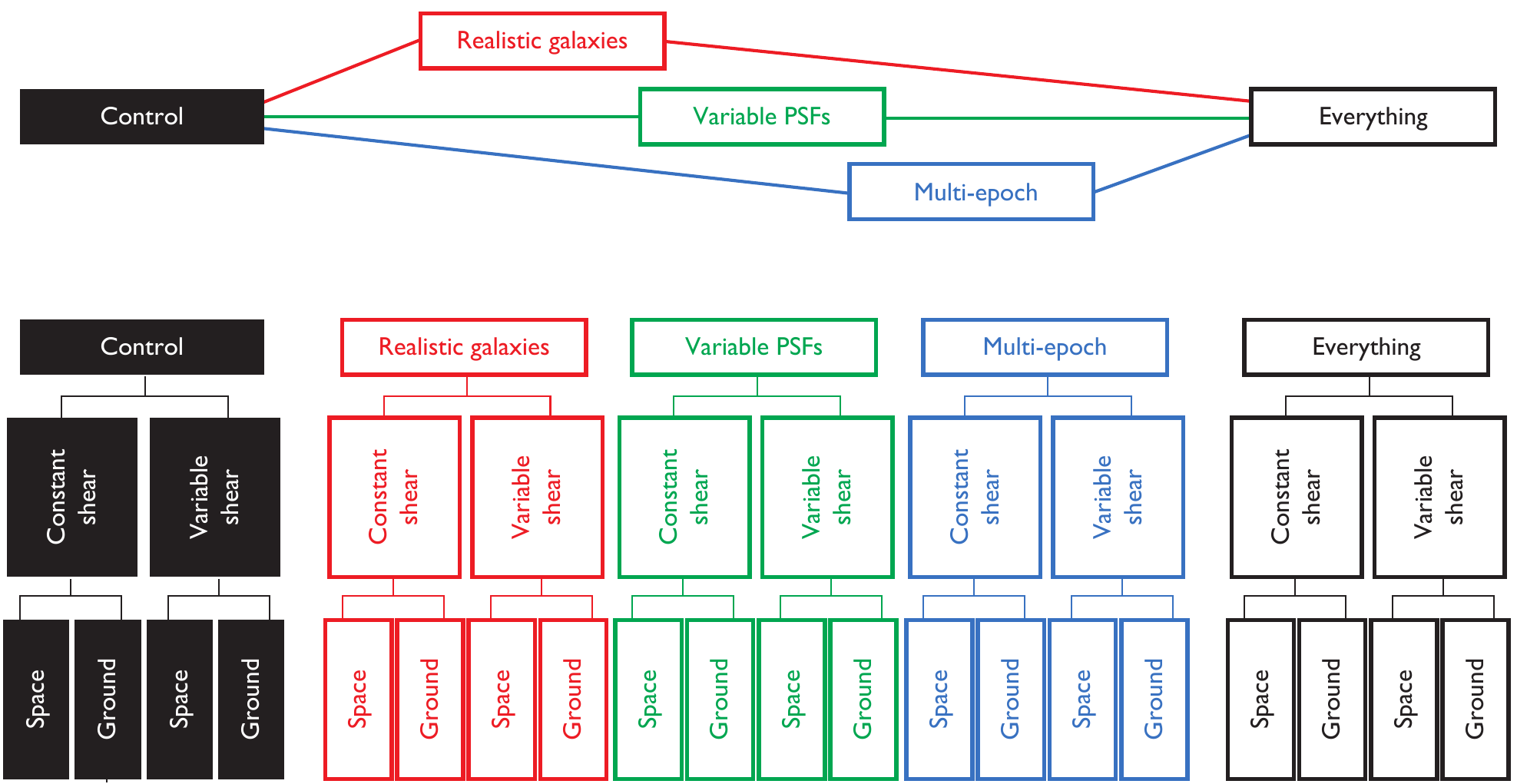}
\caption{\label{fig:branches}A schematic of the GREAT3 branch
  structure.}
\end{center}
\end{figure*}

For each experiment, there are branches with four data types: 2 shear
types (constant and variable) and 2 observation types (ground and
space).  With four data types and five experiments, we have 20 branches (Fig.~\ref{fig:branches}).

\begin{figure}
\begin{center}
\includegraphics[width=0.9\columnwidth,angle=0]{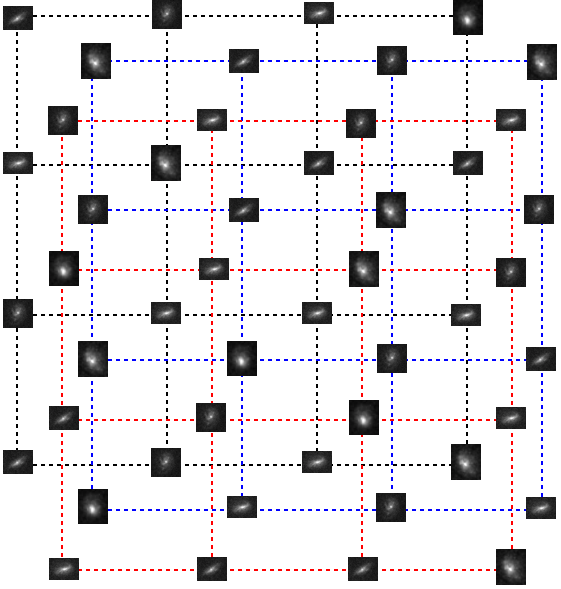}
\caption{\label{fig:subfields}An illustration of how subfields relate
  to fields, for a simple case with 3 subfields per field, each containing
  a $4\times 4$ grid of galaxies.  The image
  shows galaxies that are all part of the same field (a
  region of sky containing a particular cosmological shear field and
  PSF).  Purely for the sake of convenience, 
  rather than due to any real difference between these sets of
  galaxies, we distribute the images in subfields that consist of
  regular grids shown as dashed lines of different colors.  In our
  actual case of 20 subfields per field (for variable shear and/or
  PSF) and $100\times100$ galaxies, we have
  randomly chosen different sets of 20 offset subfield positions for each field.}
\end{center}
\end{figure}

Within
each branch, the physical setup is similar to that in previous challenges: 
there are
200 images per branch, each with a grid of $100\times100$ galaxies,
and the goal of the participants is to infer some statistic of the shear
field for each image.
The images represent $10\times 10$ degree fields.  
These images are not completely independent: each
branch of the 14 branches with variable shear and/or PSF has 10 fields representing distinct regions of the sky, but 
each field contains 20 slightly offset {\em subfields} with different
galaxies that 
sample the same shear field (in the case of variable shear).  See
Fig.~\ref{fig:subfields} for an illustration of how subfields and
fields are related.  
Thus participants must estimate the shear correlation function for
each of the 20 fields for the variable shear experiment, combining all
galaxies in all subfields when estimating the correlation function,
which can be done using
software supplied with the data (see \S\ref{subsec:overall}). 
The subfields within a field sample the same PSF pattern (see
\S\ref{subsec:sims:scales}).
For the 6 branches that have constant shear and constant PSF, the
branch contains 200 fields, each with a single subfield per field.
Thus \rev{these 6} constant shear \rev{and constant PSF} branches have 200 separate shear values. 
For the multi-epoch
simulations, each {\em epoch} of a given subfield has a different PSF;
however, a particular epoch has the same PSF for all subfields in the field.
The branches are meant to
represent the same underlying galaxy population, modulo issues that
arise when the PSF size varies (which means that galaxies
that are smaller might be simulated in one image but not
another, see \S\ref{subsec:sims:galaxies}).

\subsection{Overall information}\label{subsec:overall}

Given the challenge branch structure in Fig.~\ref{fig:branches}, we
estimate a total zipped data volume of \rev{3.2 TB}.  This figure is
dominated by the \rev{branches with variable PSFs, due to the size of
  the star fields that are provided for PSF estimation}. 
Participants may choose to submit results to any or all of those
branches at their own discretion, and likewise can download any subset
of the data that they wish.  The preferred method of getting the data
is via download from our server or its US mirror, however for a limited number of
people for whom this is not feasible, we can supply a hard drive
with the simulations.

The challenge is carried out as a competition, with a separate
leader board for each branch evaluated according to metrics described
in \S\ref{subsec:metrics}, and an overall leader board with rankings
determined based on a combination of results from the individual
leader boards as described in Appendix~\ref{subsec:leaderboard}.  There
are prizes for the first and second place winning teams
(\S\ref{subsec:rules:teams}) of the overall challenge leader board.

Detailed rules for the challenge are listed in
Appendix~\ref{sec:rules}.  Here, we summarize the online resources
related to the challenge:

\begin{itemize}
\item Webpage with leader boards, information on downloading the
  simulated data, basic information about shear conventions,
  submission format, and
  simulation file formats:\\
  \url{http://great3.projects.phys.ucl.ac.uk/leaderboard/}
\item GREAT3 web page with basic information, announcements of
  meetings: \url{http://great3challenge.info}
\item GREAT3 public code repository, which includes code that
  participants can use to automatically calculate shear correlation
  functions in a format needed for submission, a FAQ, a detailed description of the data format, and an issue
  page that participants can use to ask questions about the challenge:\\
  \url{https://github.com/barnabytprowe/great3-public}\\
  Eventually this will also include example scripts that can analyze
  all of the challenge data with some simple, existing method.
\end{itemize}

Participants may optionally sign up for a mailing list for
announcements related to the challenge data; information about this is 
available on the leader board website.  Questions
about the challenge can be sent to
\url{challenge@great3challenge.info}.

\subsection{Timeline}

\rev{A beta release of the simulations for 12 of 20 branches was released
in October, 2013, which marked the beginning of the 6-month challenge
period.  The beta period ended in late November, and the remainder of
the simulations were released in early December.  The challenge will
run until April 30, 2014, with a final meeting at the end of May 2014.}

\subsection{Evaluation of submissions}\label{subsec:metrics}

Evaluation of submissions within each branch uses metrics described
here, where the metric depends on whether the branch has
constant or variable shear.  \rev{The choice of metrics to use was
  based on experiments using simulated submissions with a grid of
  values for the multiplicative and additive shear biases.  We tested
the sensitivity of the metric to these two types of shear systematic
errors, and adopted metrics with maximal sensitivity to both.}

\subsubsection{Constant shear}\label{subsubsec:metrics:constant}

For simulations with constant shear, each \rev{field} has a particular
value of constant shear applied to all galaxies.  Participants
must submit an estimated shear for each subfield in the
branch, and the metric calculation uses those estimated shears
as follows:

Following a parametrization used in the STEP challenges and elsewhere  
\citep{2006MNRAS.368.1323H,2006MNRAS.366..101H,2007MNRAS.376...13M} we can relate biases in observed shears
$g^{\rm obs}$ to an input true shear $g^{\rm true}$ using a
linear model in each component:
\begin{equation}
g^{\rm obs}_i - g^{\rm true}_i = m_i g^{\rm true}_i + c_i
\end{equation}
where $i=1,2$ denotes the component of shear, and $m_i$ and $c_i$ are
referred to as the multiplicative and additive bias, respectively.
From the user-submitted estimates of the mean $g^{\rm obs}_i$ for each
of 200 subfields in a branch, the metric calculation begins with a
linear regression to provide estimates of $m_i$, $c_i$ given
the known true shears.  This is done in a coordinate frame rotated to
be aligned with the mean PSF ellipticity in the field, since otherwise
(with randomly oriented PSF ellipticities) the $c$ values will not
properly reflect contamination of galaxy shapes by the PSF
anisotropy.  There is a subtlety in this calculation, which is that
methods that apply weights to the galaxies will not in general give
the same weight to a galaxy and its 90-degree rotated pair (\S\ref{subsec:sims:noise}), resulting
in imperfect shape noise cancellation.  At some level, the weights  will
typically correlate with the PSF ellipticity, thus giving rise to a spurious
``$c$'' value that is due to selection bias rather than due to failure
to correct for the PSF anisotropy properly
\citep[e.g.,][]{2013MNRAS.429.2858M}.  Methods with aggressive
weighting schemes may be more susceptible to this issue.  However, as
this issue should arise in real data as well, it seems like a true
issue rather than one that occurs in simulations alone, so we do not
attempt to correct for it. 

Note that for variable PSF branches with constant lensing
shear, we are somewhat
less sensitive to additive systematics, because if the average PSF
ellipticity is zero then even in the presence of huge additive
systematics, there is no well-defined PSF direction for the field and
the additive systematics cancel out\footnote{This is not the case
  for variable shear branches, given our use of a correlation
  function-based metric.}.

Having estimated these bias parameters $m_i$, $c_i$, 
we then construct the metric for constant shear branches,
which we call $Q_{\rm c}$.  This is done by comparison of
$m_i$, $c_i$ to `target' values $m_{\rm target}$, $c_{\rm target}$.
The values of these targets are imposed by the statistical
uncertainties for upcoming weak gravitational lensing experiments: in
GREAT3 we adopt $m_{\rm target} = 2 \times 10^{-3}$ and $c_{\rm
  target} = 2 \times 10^{-4}$, motivated by the most recent estimate
of requirements for the ESA \emph{Euclid} space mission
\citep{2013MNRAS.429..661M}.  \rev{We add the $m$ and $c$ values in
quadrature with a noise term that is designed to ensure that the
scores for methods with very low $m$ and $c$ are not dominated by
noise, which can give spurious fluctuations to very high $Q_{\rm c}$.}
The constant shear branch metric is then 
defined as
\begin{equation}\label{eq:q_c}
Q_{\rm c} = \frac{2000 \times \eta_{\rm c}}{\sqrt{
 \sigma^2_{\text{min}, {\rm c}} + \displaystyle\sum_{i=+,\times} \left(
      \frac{m_i}{m_{\rm target}} \right)^2 + \left( \frac{c_i}{c_{\rm target}} \right)^2 }}.
\end{equation}
\rev{The indices $+$, $\times$ refer to the first and second
  components of the shear in the reference frame rotated to be aligned
  with the mean ellipticity of the simulated PSF in each GREAT3 image.
  We adopt values of $\sigma_{\text{min}, {\rm c}}^2 = 1~(4)$ for
  space~(ground) branches: these correspond to the typical dispersion
  in the quadrature sum of $m_i/m_{\rm target}$ and $c_i/c_{\rm
    target}$ due to pixel noise, estimated from the results of trial
  submissions to GREAT3 using the
  re-Gaussianization\footnote{In particular, we use the publicly available implementation in \textsc{GalSim} that was incorporated into an example script at\\ \url{https://github.com/barnabytprowe/great3-public}.}
  and
  \textsc{im3shape}\footnote{\url{https://bitbucket.org/joezuntz/im3shape}}
  shear estimation methods
  (\citealt{2003MNRAS.343..459H,2013MNRAS.434.1604Z}).  For methods
  displaying an $|m_i|$ or $|c_i|$ significantly greater than target
  values, the $\sigma_{\text{min}, {\rm c}}^2$ term is essentially
  irrelevant.}  This metric is normalized such that we expect a value
$Q_{\rm c} \simeq 1000$ for methods that meet our chosen targets on
$m_i$ and $c_i$. \rev{This is achieved for space branches by setting
  $\eta_{\rm c} = 1.232$, based on average scores from a suite of 1000
  simulated submissions. In the ground branches $Q_{\rm c}$ will be
  slightly lower for submissions reaching target bias levels,
  reflecting their larger $\sigma^2_{\text{min}, {\rm c}}$ due to
  greater uncertainty in individual shear estimates for ground data.
  However, $Q_{\rm c}$ scores will be consistent between space
  and ground branches where biases are significant}.  The response of
the metric to $m$ and $c$ larger than the fiducial values is shown in
Table~\ref{tab:q_c}.
\begin{table*}
\begin{centering}
\begin{tabular}{ccccccccc}
\hline
\multicolumn{4}{c}{Space simulations} & ~ & \multicolumn{4}{c}{Ground simulations} \\
Input $c_+$ & $Q_{\rm c}$ & Input $m_+$, $m_{\times}$ & $Q_{\rm c}$ & $\:\:\:$ &
Input $c_+$ & $Q_{\rm c}$ & Input $m_+$, $m_{\times}$ & $Q_{\rm c}$ \\
\hline
0.0002 & $1000$ & 0.002 & $1000$ & ~ & 0.0002 & $700$ & 0.002 & $700$ \\
0.000632 & $600$ & 0.00632 & $540$ & ~ & 0.000632 & $520$ & 0.00632 & $490$ \\
0.002 & $240$ & 0.02 & $170$ & ~ & 0.002 & $230$ & 0.02 & $170$ \\
0.00632 & $80$ & 0.0632 & $55$ & ~ & 0.00632 & $80$ & 0.0632 & $55$ \\
0.02 & $25$ & 0.2 & $17$ & ~ & 0.002 & $25$ & 0.2 & $17$ \\
\hline
\end{tabular}
\caption{Approximate average response of the constant shear metric
  $Q_{\rm c}$, defined in equation \eqref{eq:q_c}, to isotropic
  multiplicative shear bias ($m_+ = m_{\times}$) and additive shear
  bias aligned with the PSF ($c_+$). Where not otherwise specified,
  $c_+=c_{\rm target}$ and $m_+=m_{\times}=m_{\rm target}$. These
  figures were estimated from simulations of linearly biased GREAT3
  submissions, each consisting of 1000 independent realizations per
  combination of $m_+$, $m_{\times}$ and $c_+$.\label{tab:q_c}}
\end{centering}
\end{table*}

\subsubsection{Variable shear}\label{subsubsec:metrics:variable}

For simulations with variable shear, the key test is the
reconstruction of the shear correlation function.  This differs from
GREAT10, which used a metric based on reconstruction of the power
spectrum.  We adopt a correlation function-based metric because the
power spectrum-based metric requires the subtraction of shot noise
(\S\ref{subsec:sims:noise}) that contributes at all values of $k$ and
depends on the details of the shape measurement method
\citep{2012MNRAS.423.3163K}.  Subtraction of the shot noise term has
some associated uncertainty, and the real-space correlation function
is a cleaner quantity since that shot noise only contributes at zero angular separation.  Also, the correlation function-based metric is more sensitive to
additive shear systematics in the case of a constant PSF.  The
correlation function has other complications, particularly the fact
that the simplest correlation functions to calculate do not cleanly
separate into $E$ and $B$ modes, which is necessary to separate
lensing shear signals from our input $B$-mode shape noise (see
Appendix~\ref{sec:shapenoise}).  However, there is a straightforward
prescription for $E$ vs.\ $B$ mode separation from correlation
functions that does not depend on the shape measurement method, making
it a good candidate for use in a variable shear field metric, which we
now describe.

Submission of results for variable shear branches begins with
calculation of correlation functions (Appendix~\ref{sec:cf}), this
being done by the participant.  Software to calculate the correlation
function in the proper format for submission is distributed in the
GREAT3 code
repository\footnote{\url{https://github.com/barnabytprowe/great3-public}},
though participants may use their own software if they wish.  The
submission consists of estimates of a quantity called the
\emph{aperture mass dispersion}
\citep[e.g.,][]{schneider06,map_schneider}, which can be constructed
from simple $\xi_+$ and $\xi_-$ correlation function estimators, and 
allows a separation into contributions from $E$ and $B$ modes
(see Appendix~\ref{sec:cf} for details).  We label these $E$ and
$B$ mode aperture mass dispersions $M_E$ and $M_B$ in the discussion
that follows.

The submissions take the form of an estimate of $M_{E,j}$ for
each of 10 fields labelled by index $j$: this estimate is therefore
constructed using {\em all twenty} subfields in a given field.  This
choice is to provide a large dynamic range of spatial scales in the
correlation function, which helps the metric probe a greater range of
shear signals.  The $M_{E,j}$ can be estimated by the provided
software in $N_{\rm bins}$ logarithmically spaced annular bins of galaxy pair
separation $\theta_k$, where $k=1,\ldots,N_{\rm bins}$, from the smallest available angular scales in
the field to the largest.

These $M_{E,j}(\theta_k)$ are to be submitted for each field
$j=1.,\ldots,N_{\rm fields}$, where $N_{\rm fields}$ is the total number of fields in the branch.
The metric $Q_{\rm v}$ for the variable shear branches is then
constructed by comparison to the known, true value of the aperture
mass dispersion for the realization of $E$-mode shears in each field. 
These we label $M_{E, {\rm true}, j}(\theta_k)$.  \rev{The
variable shear branch metric is then calculated as}
\begin{equation}\label{eq:q_v}
Q_{\rm v} = \frac{1000 \times \eta_{\rm v}}{\sigma_{\text{min}, {\rm
      v}}^2 + \frac{1}{N_{\rm norm}}
  \displaystyle\sum_{k=1}^{N_{\rm bins}} \left|
    \displaystyle\sum_{j=1}^{N_{\rm fields}} \left[M_{E, j} (\theta_k)
    - M_{E, {\rm true}, j}(\theta_k)\right] \right| }
\end{equation}
where \rev{$N_{\rm norm} = N_{\rm fields} N_{\rm bins}$, and
  $\eta_{\rm v}$ is a normalization factor designed to yield $Q_{\rm
    v} \simeq 1000$ for a method achieving $m_1=m_2=m_{\rm target}$
  and $c_1=c_2=c_{\rm target}$ (similar to the normalization for the
  constant shear metric)}. \rev{As for the constant shear metric, we
  have added a $\sigma_{\text{min}, {\rm v}}^2$ term to ensure that
  methods that perform extremely well do not get arbitrarily high
  $Q_{\rm v}$ due to noise, but rather asymptotically approach a
  maximum $Q_{\rm v}$ value.  The order of operations (summing
  differences over fields, then taking the absolute value) is also
  intended to reduce the influence of noise. We performed suites of
  simulations using the estimates of measurement noise from
  re-Gaussianization and \textsc{im3shape} runs on variable shear branches
  in GREAT3 (see Section~\ref{subsubsec:metrics:constant}), and from
  the results of these simulations we choose $\sigma^2_{\text{min},
    {\rm v}} = 4 \times 10^{-8}~(9 \times 10^{-8})$ for space (ground)
  branches, and $\eta_{\rm v} \simeq 1.837 \times 10^{-7}$ as the
  normalization parameter, for $Q_{\rm v}$.}

\rev{For these parameter choices, Table~\ref{tab:q_v} shows the
  response of $Q_{\rm v}$ to multiplicative and additive shear
  systematics. $Q_{\rm v}$ is less sensitive than $Q_{\rm c}$,
  particularly to multiplicative biases.  This is in part due to the
  fact that shears in the variable shear branch are typically several
  times smaller than those in the constant shear branch, being drawn
  from a quasi-cosmological field (see
  Section~\ref{subsec:sims:shear}).  It is also a fact that while the
  $m_i$ and $c_i$ terms used in $Q_{\rm c}$ can be constructed from a
  linear combination of (noisy) shear estimates, any variable shear
  metric can only be estimated from second (or higher) order
  combinations of shear estimates in which the underlying signal is
  necessarily diminished relative to noise.  This feature of variable
  shear fields limits experimental sensitivity to shear biases for the
  same volume of simulation data.}
\begin{table*}
\begin{centering}
\begin{tabular}{ccccccccc}
\hline
\multicolumn{4}{c}{Space simulations} & ~ & \multicolumn{4}{c}{Ground simulations} \\
Input $c_1$, $c_2$ & $Q_{\rm v}$ & Input $m_1$, $m_2$ & $Q_{\rm v}$ & $\:\:\:$ &
Input $c_1$, $c_2$ & $Q_{\rm v}$ & Input $m_1$, $m_2$ & $Q_{\rm v}$ \\
\hline
0.0002 & $1000$ & 0.002 & $1000$ & ~ & 0.0002 & $580$ & 0.002 & $580$ \\
0.002 & $800$ & 0.02 & $900$ & ~ & 0.002 & $550$ & 0.02 & $560$ \\
0.00632 & $300$ & 0.0632 & $500$ & ~ & 0.00632 & $310$ & 0.0632 & $380$ \\
0.02 & $40$ & 0.2 & $150$ & ~ & 0.002 & $40$ & 0.2 & $125$ \\
\hline
\end{tabular}
\caption{Approximate average response of the variable shear metric
  $Q_{\rm v}$, defined in equation \eqref{eq:q_v}, to multiplicative
  shear bias ($m_1= m_2$) and constant additive shear bias
  ($c_1=c_2$).  Where not otherwise specified, $c_1=c_2=c_{\rm
    target}$ and $m_1=m_2=m_{\rm target}$. These figures were
  estimated from simulations of linearly biased GREAT3 submissions,
  each consisting of 300 independent realizations per combination of
  $m_1$, $m_2$ and $c_1$, $c_2$.  Average results for $Q_{\rm v}$ at
  $c_i$~($m_i$)~=~0.000632~(0.00632) were found to be practically
  indistinguishable from those at $c_i$~($m_i$)~=~0.0002~(0.002)
  within uncertainties.\label{tab:q_v}}
\end{centering}
\end{table*}

\rev{One could imagine other metrics, such as one that uses relative
  differences rather than absolute differences, or one that
  incorporates inverse variance weighting.  We tested these options,
  along with several others, in large sets of simulations of synthetic
  submissions: it was found that they were less sensitive than the
  $Q_{\rm v}$ metric of equation \eqref{eq:q_v} to multiplicative biases
  $m_i$ and additive systematics $c_i$ in simulated submissions.}

\section{Simulations}\label{sec:sims}

The simulations for this challenge were all produced using GalSim, 
a publicly available\footnote{\url{https://github.com/GalSim-developers/GalSim}} image simulation
tool that has been developed as a community project in part for
GREAT3, but with additional capabilities.  The software package is
fast, modular, and written in C++ and Python. 
Since it is described in detail in the documentation on the webpage
and an upcoming paper (\galsimpaper), here we
simply present evidence that GalSim can accurately simulate galaxies
with an applied shear -- see
Appendix~\ref{sec:shear-validation} for details.

\rev{The simulations are designed to provide a clean test of the
  issues raised in Sec.~\ref{sec:issues}, providing a significant
  level of realism in galaxy populations and PSF properties.  While
  the inclusion of realistically-varying cosmological shear fields in
  lensing challenges is well-established (e.g., they were used for the
  GREAT10 challenge), the galaxy populations and PSF properties that
  were used for GREAT3 represent a significant step forward in the context of a
  community challenge\footnote{\rev{Several surveys have image simulators
    that include some of these ingredients with comparable complexity
    - e.g., the LSST project has substantially {\em more} complex PSF
    models including effects not considered here - but we consider the
    inclusion of the level of realism described here in the context of
    a community challenge to have a different importance, given that
    it allows for a fair comparison of methods being adopted by any
    group rather than just those within a particular project.}}.}
Below we describe the ingredients that go into the simulations.

\subsection{Galaxies}\label{subsec:sims:galaxies}

The galaxy population in
the GREAT3 simulations is meant to represent a realistic galaxy population
in its distribution of size, magnitude, and morphology. 
\rev{This representation of the galaxy population represents a
  significant step forward for the GREAT challenges\footnote{\rev{The
    earlier STEP challenges did in fact use more realistic galaxy
    populations like those of GREAT3, rendering these galaxies first
    as simple parametric models (STEP1, \citealt{2006MNRAS.368.1323H}) and
    then shapelets (STEP2, \citealt{2007MNRAS.376...13M}, although it
    should be noted that these training
    models did not include a correction for the {\em HST} PSF
    already present in the original observations).  Our control experiment shares
    some characteristics of the STEP1 simulations though the parameter
  distributions are drawn directly from {\em HST} data, and our real
  galaxy experiment is a truly novel use of actual galaxy images in a
  community challenge for the first time.}}, bringing us 
  closer to the galaxy population observed in real surveys.} 
Here we describe how we achieve this goal.

We use data from the {\em HST} to ensure the realism of
the size and magnitude (and thus $S/N$)  
distribution, as it provides the highest resolution images available
with a uniform coverage over a reasonable area.  The way this is done depends on
whether the branch in question tests the question of
realistic galaxy morphology.  If it is, then we use actual images of
galaxies from the training sample, with a modification of the procedure
described in \cite{2012MNRAS.420.1518M} to remove the {\em HST} PSF
(unlike in that work, fully rather than partially) in Fourier space,
apply the lensing shear and magnification, convolve with the
target PSF, then return to real space and resample to the target
pixel scale.  This method generates an image
of what the galaxy would have looked like at lower resolution, provided
that the target band limit $k_\text{lim,targ}$ relates to the original
{\em HST} band limit $k_\text{lim,HST}$ via
\begin{equation}
k_\text{lim,targ} < \left(1-\sqrt{\kappa^2+\gamma^2}\right) k_\text{lim,HST}.
\end{equation}
For weak shears and convergences, the above condition is
easily satisfied by all upcoming lensing surveys, even those from space.

For the simulations that are {\em not} meant to test the effects of
realistic galaxy morphologies, we still use the {\em HST} data to
determine a distribution of galaxy properties, based on simple
parametric fits (S\'ersic) to the {\em HST} images.  We then use the
best-fitting models rather than the images themselves.  This means
that a comparison with the simulations that use
real galaxy images will \emph{directly} test the importance of underfitting bias. 

To limit the
volume of simulations needed to constrain biases in shear estimation
to levels needed for upcoming lensing surveys, we must cancel out
the dominant form of noise in lensing observations, the intrinsic shape noise
(see \S\ref{subsec:sims:noise}).  Cancellation of
shape noise requires that each measurement use all simulated galaxies
without any exclusions\rev{, which drives a minimum signal-to-noise
  cut above which there is a realistic $S/N$ distribution as described
  in \S\ref{subsec:sims:noise}.} 
Even then, noise will result in imperfect shape
noise cancellation due to chance failures to measure galaxies or
differently assigned per-galaxy weights; for more details, see
\S\ref{subsec:sims:noise}. \rev{For
multi-epoch branches, the $S/N$ values are such that the
{\em total} $S/N$ over all epochs is comparable to that in the
single-epoch branches.} 
Likewise, we exclude galaxies that are so small as to be
nearly unresolved in the simulations, since many methods will have
difficulty measuring their shapes.  \rev{The exclusion is done on a
per-subfield basis, so the galaxy populations used to simulate a
subfield with a large PSF (e.g., from the ground) will be a subset of
the population used to simulate a subfield with a smaller PSF (e.g.,
from space).} \rev{The resolution cuts do not use the 
  pre-seeing galaxy models described in Appendix~\ref{sec:galaxies}.
  Instead, they use the adaptive second moment-based resolution factor
  defined in \cite{2003MNRAS.343..459H} and precomputed using
  simulations with isotropic PSFs and no added shear, so as to ensure
  that the cuts applied on the GREAT3 simulations do not induce a 
  selection bias that correlates with the PSF ellipticity or
  cosmological shear field.}

Galaxy populations evolve with redshift,
including an increasing abundance of irregular-type morphologies and
decrease in the number of elliptical galaxies at high redshift, where
there are more young, star-forming galaxies and recent mergers (e.g.,
\citealt{2005ApJ...625..621B}).  This redshift evolution of the galaxy
population translates into a depth-dependence; for deeper
data, there are more high-redshift galaxies and therefore
more irregulars.  Thus it is relevant to ask what is the effective
depth of the simulated dataset.  Here we are limited by 
the {\em HST} dataset that we use.  If we use real images as the basis
for simulations, then the noise in 
those images also undergoes the same steps as the galaxy 
(deconvolution, shearing and magnification, convolution with the target
PSF, and resampling to the target pixel scale).  The noise in 
the result can be predicted from the
original one (since the aforementioned processes do well-defined
things to the noise) and has a direction that
correlates with both the input shear and the target PSF.  Moreover
there are generally non-negligible pixel-to-pixel correlations.  While
we can add noise that is anti-correlated and anisotropic 
to achieve isotropic, uncorrelated noise in the simulations (a process that we
call ``noise whitening'', see \galsimpaper\ for more
details), this also imposes a further limitation on the depth of the
simulated images\footnote{Technically the noise whitening procedure
  means that we treat the noise in the original images as
  part of the galaxy. This motivates us to use data for which the
  added (simulated) noise dominates
  over the existing noise in the images.}.  A simple
calculation for reasonable PSFs is that 
the effective limiting magnitude for simulated space-based data is
actually 0.6 magnitudes brighter than that in the {\em HST} training sample
given the need to whiten the substantial correlated
noise\footnote{Correlated noise is present in these images for several
reasons, including the correction for charge-transfer inefficiency
and the resampling from the instrument pixel scale to a smaller pixel
scale when combining multiple exposures.} that is
already present, whereas for simulated ground-based data it is 0.15
magnitudes brighter than the limiting magnitude in {\em HST} (because the
correlated noise gets more washed out in the simulation
process due to the larger PSF). 
However, we defer the final answer to this question of the depth of the
simulated galaxy population to
\S\ref{subsec:sims:deep}.

Details of the {\em HST} training sample and the fits to
parametrized models can be found in
Appendix~\ref{sec:galaxies}.

\subsection{Point-Spread Functions (PSFs)}\label{subsec:sims:psfs}

Here we describe the PSF model used in the challenge\rev{, several
  aspects of which are truly novel steps forward in complexity and
  realism compared to previous GREAT challenges, as described below}.  In all
simulated images, there is an optical PSF; the 
simulations of ground-based data also have an atmospheric PSF. 
The two main aspects to consider for all simulated PSFs are (1) the profile 
of the PSF and (2) the spatial variation of the PSF profiles across the focal 
plane. Both of these factors are determined by the physical 
mechanisms that give rise to the PSF. In 
\S\ref{subsubsec:sims:psf:opt} and \S\ref{subsubsec:sims:psf:atm}, we 
describe the physical basis for the optical and atmospheric PSFs in the 
``variable PSF'' experiment in this challenge. 

The simulated $10\times 10$ deg$^2$ images are much larger than the
field-of-view (FOV) of typical telescopes.  Thus, we do not 
simulate a variable PSF model for an image that size.  Instead,
for the branch with variable PSFs, we divide each image into square
tiles, and simulate a PSF model in each one of them; this underlying 
PSF model is the same for all subfields within the same field, modulo a 
position shift (which is provided; participants do not have to estimate it).
%
For the experiments with a constant PSF, we use a simplified
version of our variable PSF models, selecting a single PSF for each
$10\times 10$ deg$^2$ subfield.  For the single epoch experiments, we
make simulated images corresponding to a single deep exposure 
rather than to an idealized co-added image, because 
co-added PSF profiles can have complicated features
that would make comparison between the experiments difficult.

\subsubsection{Optical PSFs}\label{subsubsec:sims:psf:opt}
All telescopes effectively convolve the images they observe with a PSF of finite
size, due to diffraction and optical aberrations. We refer to this
contribution as the ``Optical PSF''.   In GREAT3, \rev{the first
  community challenge to incorporate an optical PSF
  model with any significant degree of realism}, 
the contributions to this PSF can be grouped into the following categories:
\begin{itemize}
\item The shape of the pupil through which light diffracts: this includes
  obscurations such as secondary mirror or instrument at prime focus,
  and obscurations due to any struts or supports for these.
\item Aberrations, which can be split into three terms: the ``design residual''
(aberrations in a perfectly realised telescope design);
``figure errors'' (aberrations due to warping or manufacturing
imperfections in mirrors); and ``alignment'' (how well mirrors are
positioned relative to each other).
\item Pointing jitter, the variation in the telescope pointing during
  an observation, which gives rise to a blurring kernel of a size and
  ellipticity that is highly design-dependent.
\item Detector response: a primary term here is the diffusion of
  charge within detectors before readout, but other (possibly
  non-linear) effects may be present, depending on the device in
  question.  Strictly speaking, any non-linear, signal dependent, or
  space variant effects cannot be represented in the simple
  convolutional model of image formation, and must be applied on top
  of that.
\end{itemize}
In all simulations with variable PSFs, we have
a fixed pupil shape and detector response across a given 
FOV for all tiles in the same image.
Typically the pupil
consists of a circular aperture upon which is superimposed a
co-centric, circular obscuration, and additional supports that are
rectangular in shape and extend radially from the central
obscuration.  

\begin{figure*}
\begin{center}
\includegraphics[width=0.33\textwidth,angle=0]{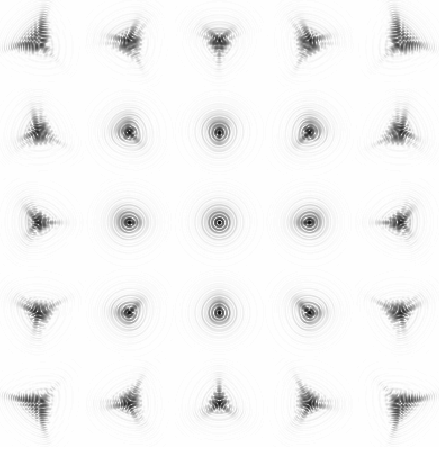}
\includegraphics[width=0.33\textwidth,angle=0]{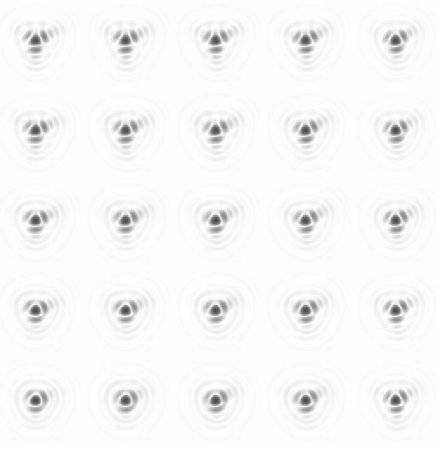}
\caption{\label{fig:optpsf}\textbf{Left:} The optical PSF (no
  atmospheric contribution) for the
  ground-based ``variable PSF'' branch at $5\times 5$ grid positions across a
simulated FOV, going all the way to the edge where aberrations are
large.  \textbf{Right:} Same as left, for the space-based 
model.  Both are shown on a logarithmic scale.  These include some (stochastic) added aberrations at a level used for the
challenge.  The space-based optical PSF model is more
constant across the field than the ground-based model because of
different assumed field-dependent aberrations.}
\end{center}
\end{figure*}

Aberrations vary across the simulated FOV, however, with a
different prescription for the ground and space-based data.
Unfortunately, we found that it is extremely difficult to make 
a realistically complex optical PSF model, including full spatial
variation, that is not instrument-specific in some way.  While our
goal is to make the GREAT3 challenge as generic as possible, we 
nevertheless had to adopt optical PSF models that are specific to
certain instruments for the ``variable PSF'' branches.  The specific
optical PSF models used for that branch are 
described in Appendix~\ref{sec:optpsf}, and some example PSF images
are in Fig.~\ref{fig:optpsf}.

For the simulations with constant PSF 
models, we adopt simple variants of the models
described in Appendix~\ref{sec:optpsf}.
For example, the space-based optical PSF model we
use for the ``constant PSF'' branches is generalized compared
to that for WFIRST-AFTA in several ways.  In particular, the size of the
basic diffraction-limited PSF is determined by the ratio of wavelength
of the light to primary mirror diameter.  
We choose a range of allowed
values for this parameter including the values for several upcoming
surveys; a range of obscuration by the secondary mirror; several
different sets of configurations for the struts.  Some additional
aberrations to represent deviations from the design residual are
included. These are evenly distributed among all the aberrations we
consider for the space-based model, and for the ground-based model
all aberrations are represented, but defocus is most important
(motivated by the realistic ground-based optical PSF
model).  As is commonly the case, the size of the additional
aberrations is a factor of several higher for the ground-based PSF
than for the space-based PSF.


\subsubsection{Atmospheric PSFs}\label{subsubsec:sims:psf:atm}

Atmospheric turbulence is the primary contributor to the PSF 
in ground-based data.  Our model for the ground-based PSF is that of a
large ($\ge 2$ m) ground-based telescope taking long exposures without
adaptive optics. \rev{As for the optical PSF model, the GREAT3
atmospheric PSF model is a step up in realism compared to previous
challenges.  The GREAT10 star challenge did include a variable PSF
model that included an atmospheric term, but the model used here is
more physically motivated due to its being }
based on a combination of high-fidelity atmospheric turbulence simulations 
and observational data. Further technical details regarding the design 
of our atmospheric PSFs can be found in
Appendix~\ref{sec:atm_psf}.

We invoke the 
LSST Image Simulator\footnote{\url{https://dev.lsstcorp.org/trac/wiki/IS_phosim}} 
\citep[PhoSim,][]{2009arXiv0912.0201L, 2010SPIE.7738E..53C, P13}, 
a high-fidelity photon ray-tracing image simulation tool, for this purpose. 
PhoSim adopts an atmospheric turbulence model similar to that used in 
the adaptive optics (AO) community \citep{1995itt..book.....R, 1998aoat.book.....H}, 
with several novel implementations to adapt 
to the wide-field nature of modern survey telescopes. The PhoSim 
atmospheric model has been shown to properly represent observational 
data \citep{P13}. Since we were concerned only with studying the effects of the 
atmosphere, we ran PhoSim in a special mode with the LSST optics removed.

First, we consider the general profile of the atmospheric PSF. To first 
order, this includes the PSF radial profile, the PSF size, and any anisotropy 
of the PSF shape. The atmospheric PSFs generated from PhoSim with
exposure times appropriate for the challenge ($> 1$ minute) has a  
radial profile that is consistent with the 
long-exposure limit atmospheric PSF predicted by a Kolmogorov model. 
The PSF profile can be written as \citep{1965JOSA...55.1427F}:
\begin{equation}
{\rm PSF} (\vec{\theta})=FT \left\{ \exp \left[ - \frac{1}{2} 6.88 \left( \frac{ \bar{\lambda} 
D |\vec{f}| }{r_{0}} \right)^{5/3} \right] \right\}
\end{equation}
\noindent where ``$FT$'' represents a Fourier transform between angular 
position $\vec{\theta}$ and spatial frequency $\vec{f}$, $\bar{\lambda}$ is the 
average wavelength, $D$ is the aperture size, and $r_{0}$ is the Fried parameter. 

Given a survey design, the Kolmogorov PSF takes one parameter, 
$r_{0}$, which can be rephrased in terms of the commonly-used ``atmospheric 
seeing'', defined as the full-width at half maximum (FWHM) of the atmospheric PSF. 
We adopt a distribution of FWHM values estimated at the summit of 
Mauna Kea in one optical filter ($R$, $\langle
\lambda\rangle\sim 651$~nm) at 
zenith\footnote{Figure 1 from\\
\url{http://www2.keck.hawaii.edu/optics/ScienceCase/TechSciInstrmnts/Products_SeeingVarMaunaKea.pdf}.}. 
The quartiles of
this distribution are 0.49, 0.62, and 0.78 arcsec; the mean value is
0.66 arcsec. For a single exposure, 
we draw a value of atmospheric PSF FWHM from this distribution.  These
are not purely random; since single-epoch experiments only have 10
different PSFs in the entire branch, and we want to properly cover
this distribution, we draw randomly from within percentiles, i.e., one
field has a PSF drawn randomly from below the tenth percentile in the
distribution, another is from the tenth to twentieth percentile, and
so on. 
Finally, these Kolmogorov PSFs are assigned an ellipticity
(Eq.~\ref{eq:ellipticity})
to represent the small anisotropy in the atmospheric PSF. The ellipticity 
values are based on a large number of PhoSim simulations.

\begin{figure}
\begin{center}
\includegraphics[width=\columnwidth,angle=0]{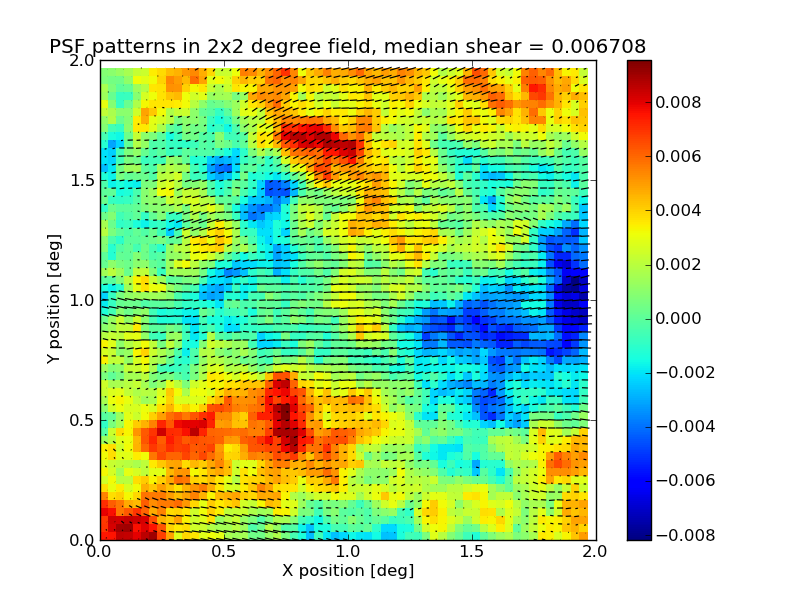}
\caption{\label{fig:atmos_example}A single random realization of an
  atmospheric PSF anisotropy pattern in a $2\times 2$ deg$^2$ field, for a 2 minute
  exposure at a 4-meter telescope.  The plot title gives the
  median PSF shear.  The color scale indicates the fractional change
  in size of the atmospheric PSF as a function of position.}
\end{center}
\end{figure}

Next, we consider the spatial variation of the PSF model parameters 
(size and ellipticity), 
quantified by a 2-point correlation function.  We find 
that the spatial variation of atmospheric PSF parameters in PhoSim can be well described by a 
functional form with two parameters. We generate the 
spatially varying PSF parameters as a Gaussian random field that 
corresponds to this correlation function, with the two model
parameters allowed to vary in a reasonable range. An example of an ellipticity
field and the spatial variation of PSF size generated from this procedure (described in more detail in
Appendix~\ref{sec:atm_psf}) is shown in Fig.~\ref{fig:atmos_example}.
\begin{figure}
\begin{center}
\includegraphics[width=0.95\columnwidth,angle=0]{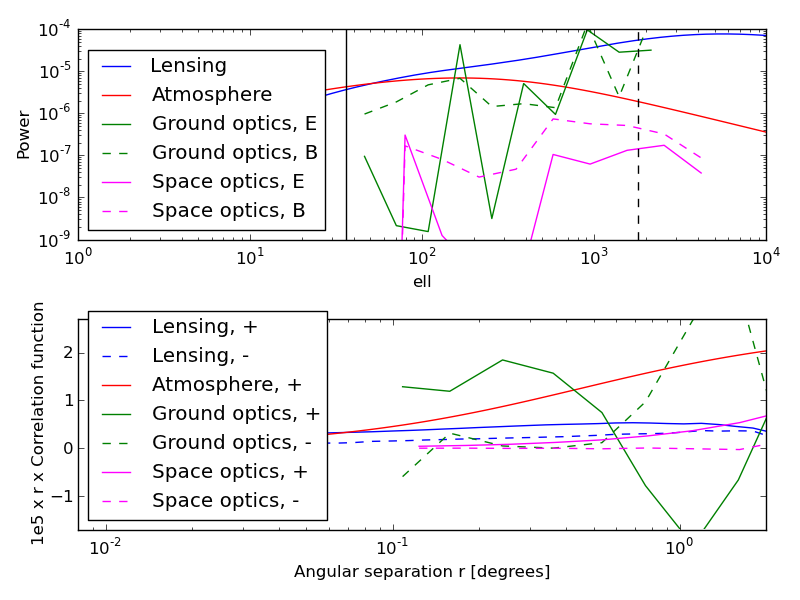}
\caption{\label{fig:ps_cf} \textbf{Top:} Dimensionless power of lensing
  shears, atmospheric PSF shears, and optical PSF shears (where the
  latter is computed by tiling several adjacent pointings for our 
  ground-based optical PSF model -- without additional tilt,
  misalignment, or defocus -- after convolving with a circular
  0.7\arcsec\ Kolmogorov blur for the ground-based model).  For  
  lensing, the power spectrum that is shown is $E$ mode,
  and the $B$ mode power spectrum is zero.  For the atmospheric PSF,
  the power spectrum shown is the same for $E$ and $B$.  The solid
  black line shows the minimum accessible $\ell$ value given the size
  of our images.  The dashed black line shows the maximum $\ell$ value
  given the grid spacing on our images; however, since 
  multiple images sample the same shear field, the true maximum
  $\ell$ is actually larger.  \textbf{Bottom:} Correlation functions
  for the three cases shown above, after multiplying by separation on
  the sky in degrees.  Here we show $\xi_+$ and $\xi_-$,
  but the latter is identically zero for our atmospheric PSF
  model.}
\end{center}
\end{figure}

Our choice to use a sheared Kolmogorov profile (without any
higher-order distortions) is a simplification 
compared to reality, \rev{but inspection of the PhoSim simulation images
showed that, for reasonable
exposure times and telescope sizes, it is correct to a good
approximation.
Hence we consider our prescription to be realistically complex enough
for an interesting and relevant test, and note that the modelling of
atmosphere-driven variations in PSF size and ellipticity is a significant
enhancement in realism compared to previous projects of this kind.}

Fig.~\ref{fig:ps_cf} shows a comparison between the power spectrum and
correlation functions of the lensing shears, the atmospheric PSF
anisotropies, and the ellipticity of the optical space- and ground-based PSF
model (in the latter case, after convolving with a circular,
typical-sized atmospheric PSF).  Here we have omitted the aberrations
other than the design residual to get an idealized version of the
results for the optical PSF model. This plot 
shows the most important scales for the various systematics
compared to the weak lensing shears.  For example, we see that the
lensing power spectrum is below that of the atmospheric PSF
anisotropies on large scales (small $\ell$). 
However, for nearly 
all relevant scales on our grid, the atmospheric PSF anisotropy
correlation function is greater than
that of the lensing shear.  The optical PSF anisotropy is primarily
relevant on small scales (small angular separation or high $\ell$).
However, because of the tiling of multiple fields of view, it can be
important on large scales, particularly for the space-based model.

\subsubsection{Star fields}

In the constant PSF experiments, we provide several noiseless images of the
PSF for each image, at the same resolution as the galaxy images.  One
of those images is centered within a postage stamp; the others are randomly
offset by some amount to be determined by participants, who may use
them if they wish to recover information about the PSF on sub-pixel
scales.  For the space-based images with single exposures, the PSF is
Nyquist sampled, and hence those offset images carry no additional
information.

In the variable PSF
experiment, we provide star fields that can be used for PSF
estimation, one star field per subfield (however the 20 subfields in a
field have the same underlying PSF).  If our 20 slightly offset
grids of $100\times 100$ galaxies that belong to the same field 
cover the same $10\times 10$ deg$^2$ area of the sky, and we
want to simulate a realistic stellar density\footnote{Considered as an 
average over stellar densities for different galactic latitudes at reasonable 
galactic altitudes.} of $2$~arcmin$^{-2}$ down to $S/N=50$ for an image that goes to $r$
magnitude of $25$, that means each star field has $\sim 1.3\times10^4$
randomly-located stars\footnote{However, a small exclusion radius is
  placed around each one to avoid blending effects.}.  The magnitude 
distribution for the star fields is motivated by the model in \citet{2008ApJ...673..864J}.
Some methods use only very high
$S/N$ stars resulting in $\sim 1$~arcmin$^{-2}$, but those that can go
to lower $S/N$ will find a higher usable stellar density and may be
able to better trace the small-scale fluctuations in the PSF.  For the
experiment containing all effects, each epoch will have its own star
field for PSF estimation, since the PSF varies per epoch. 

\subsection{Noise model}\label{subsec:sims:noise}

In a weak lensing measurement, two important sources of noise
are ``shape noise'' (the intrinsic,
randomly-oriented 
galaxy shapes that we must average out to measure the small,
coherent lensing shears), and ``shape measurement error'', the noise
in individual galaxy shape measurements due to the noise in each pixel.  For
typical galaxy populations, the shape noise dominates over measurement
error for all but the very lowest signal-to-noise galaxies, 
 where the two might become comparable.  Together these sources of
 error are often called ``shot noise''.

Previous challenges have incorporated schemes to cancel out the shape
noise, thus substantially reducing the volume of simulations needed to
test shear measurement methods very accurately from of order 10 TB to
$\sim 1$ TB.  Shape noise cancellation is imperfect due to
measurement error, but it is still reasonably effective down to
observations with $S/N\sim 20$ (for $S/N$ defined in
Eq.~\ref{eq:sndef} \rev{below; this is an idealized $S/N$ estimate, defined
  before application of noise to the images, with an optimal weight
  that is therefore unachievable in any real measurement}).  For
galaxies with lower $S/N$, the 
noise typically leads to a substantial measurement failure rate that
renders shape noise cancellation very ineffective.   Shape noise
cancellation also eliminates some 
selection biases.

Given the limitation imposed by our desire to keep the
simulation volume under control, in GREAT3 we employ shape noise
cancellation, with a lower limit on the \rev{(optimal)} galaxy $S/N$
of 20\rev{, though we will discuss the effective $S/N$ limit with a
  more realistic estimator later in this section}.  In
the constant-shear simulations, shape noise cancellation is carried
out by having the same galaxy included twice, with
orientations rotated by 90 degrees before shearing and PSF
convolution \citep{2007MNRAS.376...13M}.  
Given the typical $S/N$ and intrinsic ellipticity distribution for the
galaxies in our parent sample, the shape noise cancellation scheme
reduces the errors on measured shears by a factor of 3 (equivalent to
$9\times$ simulation volume).  We 
have tested the effect of completely random galaxy omissions (e.g.,
due to convergence failure for some shape measurement method), and
found that for simulated data with typical $S/N$, the errors on the
measured shear increase from the optimal case (perfect shape noise
cancellation) by 8\%, 30\% and 50\% for the case of 5\%,
10\%, and 20\% missing galaxies, respectively.  This is still
well below the 200\% increase that corresponds to no shape noise
cancellation, so even for a significant random failure rate the
errors on the shear (and therefore metric) increase, but not so much 
that the results become useless.

In the variable-shear simulations, as for GREAT10, the 
lensing shear is entirely $E$-mode power (as in reality) and 
shape noise is only in the $B$-mode (this is completely
unrealistic, but useful).  This task is more complicated
than for GREAT10 given our use of a real galaxy population;
see Appendix~\ref{sec:shapenoise} for a description of how we carry
out shape noise cancellation.

It thus remains to describe our model for pixel noise.  In real data,
pixel noise is largely Poisson (since the CCDs are counting photons)
but with a small Gaussian component from detector read noise.  In many
datasets, Poisson noise is dominated by the sky rather than the
objects, except for very bright ones that constitute a small fraction
of the objects used for shape measurement.  Moreover, the sky level is
often high enough that its Poisson noise is essentially Gaussian.  We
therefore employ a Gaussian noise model only\rev{, corresponding to a
  single constant variance and a mean of zero.  The variance, which
  must be estimated directly by those participants whose method
  requires an estimate of noise variance, is constant throughout a
  single subfield image, but can vary for different subfields.  While
  the use of effectively sky noise only (no noise from the objects) is
  a simplification, it is not a very problematic one for the galaxies:
  as shown in, e.g., \cite{2004MNRAS.353..529H} and
  \cite{2012MNRAS.427.2711K} for two quite different methods of shear
  estimation, the realm in which noise bias is most problematic is for
  galaxies with $S/N\lesssim 30$, corresponding to galaxies for which
  sky noise dominates over the noise in the galaxy flux.  Thus this
  simplification is acceptable, and has the added benefit of
  simplifying other aspects of the simulations (e.g.,  we do not have to explore different values of gain, and we
have a relatively simple $S/N$ estimator for the galaxies).}
  
Our definition of galaxy $S/N$, which we use to decide which galaxies
go into the simulations, is the same as for GREAT08
\citep{2010MNRAS.405.2044B}.  We define the signal as a weighted
integral of the flux,
\begin{equation}\label{eq:signal}
S = \frac{\sum W({\bf x}) I({\bf x})}{\sum W({\bf x})}
\end{equation}
and its variance is
\begin{equation}\label{eq:signalvar}
\text{Var}(S) = \frac{\sum W^2({\bf x}) \text{Var}(I({\bf x}))}{(\sum W({\bf x}))^2}.
\end{equation}
In the limit that the sky background dominates, $\text{Var}(I({\bf x}))$
is a constant, so we can just call it $\text{Var}(I({\bf x}))=\sigma^2$,
the pixel noise variance (\rev{this simplification depends on our
  adopted noise model, and would not be more generally valid}) .  We adopt a matched filter
for $W$, i.e., $W({\bf x})=I({\bf x})$.  Putting those assumptions into
Eqs.~\ref{eq:signal} and \ref{eq:signalvar} gives
\begin{equation}\label{eq:sndef}
S/N = \frac{\sqrt{\sum I^2({\bf x})}}{\sigma}.
\end{equation}
While we do not have noise-free images for the real
  galaxies (for calculating the sum over squared intensities), we can
  use the model fits to the galaxy images as noise-free images for this
  purpose.

\rev{It is important to remember that this optimal $S/N>20$ constraint
  does not correspond to a $S/N>20$ cut that would be applied
  using some typical $S/N$ estimator on the real data.  For example,
  {\sc sextractor} \citep{1996A&AS..117..393B} analysis of the
  simulated GREAT3 images gives a distribution of $S/N$ values that
  has one-sided 99\% and 95\% lower limits of 10.0 and 12.0 (ground)
  or 11.7 and 13.2 (space) for single-epoch simulations.  Hence the
  galaxy $S/N$ distribution that is being simulated is in fact comparable to
that in samples that are used for real weak lensing analyses,
including the potential for significant noise bias
\citep{2004MNRAS.353..529H,2012MNRAS.427.2711K,2012MNRAS.423.3163K,2012MNRAS.424.2757M,2012MNRAS.425.1951R}.}

This $S/N$ definition is also used for the stars in the star fields
when defining a $S/N$ limit.  For stars, the assumption that background
dominates is not very realistic.

  Also, as described in \S\ref{subsec:sims:galaxies}, the
  original training data in the ``realistic galaxy'' branches has
  noise in it already, so we only add enough noise to ensure that the
  resulting noise correlation function is the target one, i.e.,
  Gaussian noise with $\sigma$ defined by Eq.~\ref{eq:sndef}, without 
  pixel-to-pixel correlations.

Many image
processing steps that are carried out on real data, especially
from space telescopes, can lead to correlated noise, due to stacking of multiple exposures.
For simplicity we include only uncorrelated noise in
GREAT3. 

\subsection{Image rendering}\label{subsec:sims:rendering}

GalSim provides two primary methods of rendering images of a galaxy that has
been sheared/magnified and convolved with a PSF: via discrete Fourier
transform (DFT), and via photon-shooting.  The latter method was used by the
software for the GREAT08 and GREAT10 challenges, and involves
representing shears, magnifications, and convolutions as offsets of
photons that were originally drawn according to the light distribution
of the intrinsic galaxy profiles.

However, for the GREAT3 challenge, we have adopted DFT as our method
of image rendering, for the following reason: to use real galaxy
images as the basis for our simulations
(\S\ref{subsec:sims:galaxies}), we need to remove the original PSF
from the {\em HST} images.  There is no way to represent
deconvolution in a photon-shooting approach, and so
for consistency, all branches of the GREAT3 challenge (even those that
use parametric galaxies) are generated using DFT.

However, since the two methods use different approximations, our tests
of the image rendering in Appendix~\ref{sec:shear-validation} include
a comparison of DFT versus photon-shooting as a way to validate the
results.

\subsection{Constant versus variable shear}\label{subsec:sims:shear}

The challenge consists of two shear types. 
Half of the challenge branches contain images with a single
constant value of shear for the image, and the other half contain
images that have a variable shear field.  The justification for
this division is that some lensing measurements, like galaxy-galaxy
lensing, can be carried out by averaging some roughly constant shear
value within annuli around lens object(s), whereas cosmic shear
measurements involve estimating the variable shear field caused by
large-scale structure.  Both types of measurements are
scientifically useful.  Additionally, some shear estimation methods
may work better in one regime than the other; stacking methods
\citep[e.g., ][]{2009MNRAS.398..471L,2010MNRAS.405.2044B} are simplest to interpret in the constant
shear regime, whereas methods that assume something about the statistical
isotropy of the galaxy shape distribution may fail in a constant shear
field.

For the constant shear branches, simulations have a single constant
value of shear drawn randomly from a hidden distribution in $|g|$ with
some minimum and maximum value, with purely random position angles.

In the variable shear branches, we start with a 
shear power spectrum with reasonable shape for a typical cosmology,
and with slightly high amplitude in order to increase sensitivity of
$Q_{\rm v}$ to multiplicative biases.  Then, we include 
a nuisance function that gives scale-dependent 
modifications of order $\sim 10$\% on the range of scales 
traced by our grid of galaxies.  In a single dimension, the angular
grid extent of $L=10$ degrees ($L_\text{rad}=\pi/18$ radians)  means that the minimum 
relevant $\ell$ value is $\ell_{\rm min}=2\pi/L_\text{rad}=36$.

This shear power spectrum is given as input to GalSim,
which uses it to generate a realization of a Gaussian random shear
field, and also generates
self-consistent convergences.  The resulting values of shear $\gamma$
and convergence $\kappa$ are used to shear the galaxy according to the
reduced shear $g$ (\S\ref{sec:physics}) and to magnify the
galaxy according to the magnification\footnote{This procedure only includes
changes in observed galaxy sizes and fluxes; it does not include the other important effect of
magnification (the modification of the number density of objects due
to the change in solid angles and the fact that galaxies get scattered
across cuts in flux and apparent size by the magnification process).}
$\mu = [(1-\kappa)^2 - \gamma^2]^{-1}$.

The GalSim ``lensing engine'' that
carries out this process works in the flat-sky limit.  It uses Fourier transforms, 
with a Fourier-space grid that is of equal size to the real-space
grid, and hence the power is assumed to be zero for $|k|<2\pi/L$ and
$|k|>\sqrt{2}\pi/\Delta x$ (see Appendix \ref{sec:finitesim} for
details).  This artificially
reduces shear correlations on large scales by a significant amount
compared to those in a realistic shear field (\galsimpaper).  To
address this limitation, we use 
an extended real-space grid for calculating shears, which lowers
the minimum $k$ represented in the power spectrum and preserves the
shear correlations on scales corresponding to our box size.


Because of various effects that modify the power spectrum at
levels up to a few percent (e.g., reduced shear, random chance in a
single realization of the shear field, flat sky approach, and the
limited Fourier space grid used to generate the shears), we 
do not compare submitted shear correlation functions with the {\em input}
ideal ones, but rather with correlation functions that we estimate
using the true reduced shears output from GalSim before they are actually applied
to the galaxies.

\subsection{A note about physical scales}\label{subsec:sims:scales}

In real images, galaxies may be quite close together (given typical
number densities of $\sim 20$ arcmin$^{-2}$), yet in the case of
variable shear fields we usually only estimate shear correlations for
galaxies that are significantly farther apart than the average
separation between galaxies.  This fact has motivated GREAT10 and now
GREAT3 to consider galaxy grids that are $10\times 10$ deg$^2$ with
$100\times 100$ galaxies, not spending time simulating galaxies that
are very close together.

However, for variable PSFs, much of the interesting PSF
variation happens on smaller scales than the $0.1$ deg
grid spacing.  This has
motivated us to make each variable PSF branch contain ten fields of 20 subfields that
sample the same cosmological shear field and PSF field, thus sampling
the PSF field more densely than the cosmological shear field. This
also aids us in the calculation of the metric, \S\ref{subsec:metrics}.

\subsection{Space versus ground}

Much of the difference between space-based and ground-based data comes
from the different PSFs, as described in
\S\ref{subsec:sims:psfs}.  The PSFs in space-based data are smaller 
and more
stable over time than ground-based PSFs.  However, there is an 
additional difference that is included in the GREAT3 challenge,
related to the sampling of the images.

Data from existing optical space telescopes, as well as planned future
telescopes, are typically undersampled due to the relatively large
pixel scale compared to the PSF size.  Sub-pixel dithers are used to
recover Nyquist sampled data after combining multiple exposures.
However, since the combined image typically has a smaller pixel scale
than the original image, the combined image has other features such as
correlated noise (and depending on how the image combination is
carried out, there might be some aliasing - see, e.g.,
\citealt{2011ApJ...741...46R}).

In the control, realistic galaxy, and variable PSF experiments, the
simulated data do not have multiple exposures.  Thus, if we are
simulating space-based data, we need some way to have that single
exposure be Nyquist sampled.  Our choices are (a) to simulate some
realistic co-add over multiple single exposures, including effects
like correlated noise, or (b) to simulate what would happen if our
detectors had smaller pixels that allowed them to be Nyquist sampled
while also having uncorrelated noise.  We opt for choice (b).  In the
multi-epoch and full experiments, the individual exposures have
pixel scales that are larger and hence not Nyquist sampled until all
exposures are combined.  In practice, we use pixel scales of
0.1\arcsec\ and 0.05\arcsec\ for simulated multi-epoch and
single-epoch data, respectively.

In contrast, ground-based data is rarely undersampled, and we adopt a
single pixel scale of 0.2\arcsec\ for the simulated ground-based data,
regardless of whether it is single- or multi-epoch.

\subsection{Deeper data}\label{subsec:sims:deep}

Many lensing surveys that are planned for the near- and far-future are
designed with both ``deep'' and ``wide'' components.  The ``deep''
components are typically a small subset (few percent) of the area of
the ``wide'' component, but include enough observations to increase
the $S/N$ by a factor of several.  These 
deep fields enable training methods to 
learn something about galaxy populations,
which can then be used when interpreting the data in the (more
cosmologically-interesting) wide survey.

To facilitate tests of such training methods, the GREAT3 challenge
has additional simulations for each branch (corresponding to 2.5\% of
the volume of that branch, i.e., 5 images) that represent data that
are one magnitude deeper (2.5 times higher $S/N$) than the rest
of that branch, but are otherwise drawn from the same underlying galaxy
population. \rev{The shears and PSFs in the deeper images are determined
according to the same rules used for the rest of the branch.}  The
deeper data are not to be used to estimate shears, and results for
them should not be submitted; they are purely for use as a training
dataset. 

In a real dataset, the deep survey would include a large fraction
of galaxies that are not even detected in the wide survey.  However,
since we do not want most of the galaxies in the
GREAT3 deep data to be useless, we only simulate the ones that
would be observed in the rest of the GREAT3 challenge \rev{with $S/N$
  above our limiting value}. 
The galaxies that are simulated in the deep dataset 
still have resolution cuts imposed according to the PSF size in the
deep dataset.  The
population is therefore identical, but with $S/N>50$ in the deep
dataset, which means that the effective deep data fraction is actually
$5$--$7.5$\% rather than 2.5\%
This volume of deep data is actually relatively high compared to 
many planned surveys, but since the amount of deep data needed is
still an open question, a test with this amount of deep data is quite
useful.

Our interest in simulating a galaxy population in the challenge that
goes to $I<25$ with limiting $S/N=20$, but to also have a subset of
simulations in which the effective $S/N$ limit {\em for the same
  population} is $S/N=50$, poses a difficulty for our training
dataset.  The depth of our {\em HST} training dataset
(\S\ref{subsec:sims:galaxies} and Appendix~\ref{sec:galaxies}) is such
that at $I=25$, the images we observe have $S/N$ below $50$.  We are
forced to conclude that if we wish to have a limiting $S/N$ of $50$
in the deep simulations for a magnitude-limited parent sample from
{\em HST}, we must use $I<23.5$.  This is relatively shallow compared
to 
many extant and future lensing surveys, and hence somewhat
undesirable.  To ameliorate this issue, we developed a simple
procedure to use the $I<23.5$ sample to mimic the observed properties
of an $I<25$ sample by simple changes in flux and radius; a
description and tests of this procedure are in
Appendix~\ref{subsec:fake}.  This procedure does not
preserve the {\em intrinsic} properties of the galaxies such as their
redshift distribution, luminosity distribution, or intrinsic size
distribution.  However, it allows us to use the $I<23.5$ sample to
match the quantities that dictate the shear systematics 
for an $I<25$ sample - namely the $S/N$, {\em observed} size, and
observed morphology.  This also helps address the concern raised in
\S\ref{subsec:sims:galaxies} that noise in the original {\em HST} images is
treated as part of the galaxy; with the scheme described here, the
added noise dominates over the noise that was already present for all galaxies.

\section{Simplifications}\label{sec:future}

As described in \S\ref{subsec:philosophy}, our goal of making
simulations to test particular effects has led to some
simplifications.  Here we briefly mention several classes of problems
that are left for future work:

\begin{itemize}
\item \textbf{Non-gridded galaxies:} Since our simulations include
galaxies on grids at known locations, we do not test for issues due to
blends (overlapping galaxy profiles).
\item \textbf{Selection biases:}
Imposition of selection criteria that might lead to the probability of
a galaxy being selected to correlate
with the shear or PSF direction are not tested in this scheme.
However, if the weights used for a particular galaxy depend on the
shear or PSF, then that form of selection bias will show up in the
challenge results.
\item \textbf{Wavelength-dependent effects:} Real PSFs are 
color-dependent at some low level.  Since star and galaxy SEDs are not
the same, this results in a different effective PSF for galaxies as
for stars.  Further complications arise due to color gradients
within galaxies.  None of these effects are included in GREAT3, but
may be quite important for upcoming lensing surveys
\citep{2013MNRAS.432.2385S}.
\item \textbf{Instrument/detector specific effects:} There are a whole
host of instrument and detector effects that are not included in
GREAT3, for example cosmic ray hits, saturation, 
bad pixels or columns, scattered light, charge transfer inefficiency,
and distortion.  Because these effects are manifested in different 
instruments in different ways, it is difficult to simulate them
in a generic context.
\item \textbf{Star/galaxy separation:} In a realistic data analysis,
  it is necessary to determine from the data itself which objects are
  galaxies (to be used for shear estimation) and which are stars (to
  be used for PSF modeling).  Cross-contamination between the samples
  in either direction can cause biases in shear estimation.  Since the
  GREAT3 challenge provides galaxy fields for shear estimation, and
  star fields for PSF estimation, star/galaxy separation is not
  explicitly tested by this challenge.
\item \textbf{Background estimation:} The images in the
  GREAT3 challenge have essentially had the sky background level
  subtracted.  In practice, the sky background level is unknown
  and spatially varying; incorrect removal (especially 
  contamination by the extended light of galaxies) can lead to unsubtracted sky
  level gradients that mimic shear signals.
\item \textbf{More complex noise model:} In real images, the noise may
  be more complex than the simple model adopted here
  (\S\ref{subsec:sims:noise}).  Aside from the issue of
  spatially-varying sky background and Poisson noise from the galaxy
  flux, some steps in image processing can induce correlations in
  noise levels between pixels.
\item \textbf{Redshift-dependent effects:} The GREAT3 challenge does
  not include redshift-dependent shears or allow for estimation of a
  redshift-dependent shear calibration.
\item \textbf{Flexion:} The GREAT3 challenge assumes the shear is
  constant across each galaxy, so it does not include higher-order
  distortions such as flexion.
\end{itemize}

\section*{Acknowledgments}

The authors of this work benefited greatly from discussions with
Christopher Hirata, Gary Bernstein, Lance Miller, and Erin Sheldon;
the WFIRST project office, including David Content; the Euclid
Consortium; and the LSST imSim team, including En-Hsin Peng; and Peter
Freeman.  We thank the PASCAL-2 network for its sponsorship of the
challenge.  This work was supported in part by the National Science
Foundation under Grant No. PHYS-1066293 and the hospitality of the
Aspen Center for Physics.

This project was supported in part by NASA via the Strategic
University Research Partnership (SURP) Program of the Jet Propulsion
Laboratory, California Institute of Technology; and by the IST
Programme of the European Community, under the PASCAL2 Network of
Excellence, IST-2007-216886.  This article only reflects the authors'
views.

RM was supported in part by program HST-AR-12857.01-A, provided by
NASA through a grant from the Space Telescope Science Institute, which
is operated by the Association of Universities for Research in
Astronomy, Incorporated, under NASA contract NAS5-26555.  BR and SB
acknowledge support from the European Research Council in the form of
a Starting Grant with number 240672. HM acknowledges support from JSPS
Postdoctoral Fellowships for Research Abroad. CH acknowledges support
from the European Research Council under the EC FP7 grant number
240185.  FC and MG are supported by the Swiss National Science Foundation (SNSF).

\bibliography{handbook}

\appendix

\section{Existing approaches to shear measurement}\label{sec:shearmeas}

Initially, the field of weak lensing was dominated by methods that involved
applying a correction to the weighted second moments of the galaxy
image to account for smearing by the PSF.  These early methods include
KSB \citep{1995ApJ...449..460K} and implicitly make unrealizable
assumptions about the nature of the galaxy and PSF: that they have
concentric isophotes \citep{2007MNRAS.376...13M},
and small intrinsic ellipticities \citep{2011MNRAS.410.2156V}.  Since
then, the weak lensing community has made significant progress in developing
additional PSF-correction methods.
Like KSB, some of those methods also start with measuring moments of
the galaxy and PSF, with some prescription for correcting the former
to account for the latter (e.g.,
\citealt{2000ApJ...537..555K,2000ApJ...536...79R,
  2003MNRAS.343..459H}).

Other methods are based on
forward modeling of the intrinsic galaxy profile, 
including some methods that carry out $<10$ parameter fits for an
astrophysically-motivated galaxy model (e.g.,
\citealt{2013MNRAS.434.1604Z} and \citealt{2013MNRAS.429.2858M}), and
others that decompose the galaxy images into an orthonormal basis set
(e.g., \citealt{2002AJ....123..583B}, \citealt{2005MNRAS.363..197M},
and \citealt{2009MNRAS.396.1211N}),
requiring many more parameters but also allowing a lot more
flexibility for describing complex galaxies.  Additionally,
several methods have gone in newer directions such as Fourier-space
approaches and non-parametric methods
\citep{2010MNRAS.406.2793B,2011MNRAS.414.1047Z,2013arXiv1304.1843B}.
For measurements of constant shears, image stacking methods (which were highly successful in the GREAT08
Challenge, \citealt{2009MNRAS.398..471L,2010MNRAS.405.2044B}), have a clear potential
application.
 
Several studies have assessed the limitations of previous
methods and devised ways of compensating for them (e.g., 
\citealt{2002AJ....123..583B}, \citealt{2003MNRAS.343..459H},
\citealt{2010AJ....140..870B}, and
\citealt{2011MNRAS.410.2156V}).  A relatively new development in the
GREAT10 challenge \citep{2012MNRAS.423.3163K} was the introduction of
several methods using techniques from machine learning
and computer science, such as the use of training methods (neural network and
lookup table approaches, e.g., \citealt{2010ApJ...720..639G}).  

Because of the wealth of information
about these methods in the literature and in the summaries of the
GREAT08 and GREAT10 challenges, we refer the interested reader
to the relevant papers and references therein for more details on modern shape measurement methods.


\section{Estimating the shear correlation function}\label{sec:cf}

For the variable shear branches, we cannot use the average shear as a
useful metric to decide whether a given set of shear measurements match
the input shear field.  In fact, the average input shear is zero by 
construction.  Instead, we use the two-point correlation function of
the shear field.  This statistic is commonly used in weak lensing 
cosmic shear studies as the lowest order description of the shear
field in a given patch of sky.  For a Gaussian field, as is used in this
challenge, it encapsulates all of the measurable information about the 
shear field.\footnote{
The actual cosmic shear field of the universe is not Gaussian,
so higher order statistics such as three-point correlation functions and
shear peak statistics, among others, are also used to characterize the
non-Gaussian features in the shear field.}

As the name implies, the ``two-point'' correlation function involves an 
average over all pairs of two shear measurements.  The math is simplest
if we treat the shears as complex numbers, $g = g_1 + {\rm i} g_2$. 
Because of the complex nature of shear, there are actually two shear
correlation functions, $\xi_+$ and $\xi_-$, defined as follows:
\begin{align}
\xi_+(r) &= \langle g({\bf x}) g^*({\bf x} + {\bf r}) \rangle \label{eq:xip} \\
\xi_-(r) &= \langle g({\bf x}) g({\bf x} + {\bf r}) e^{-4{\rm i}\alpha} \rangle \label{eq:xim}
\end{align}
where the averages are over all pairs of measured shear values, $\alpha$ is the polar angle of ${\bf r}$ and ${}^*$ indicates complex conjugation.

Both $\xi_+$ and $\xi_-$ are complex-valued by construction, but 
they are both effectively real in practice.  In fact, $\xi_+$ is identically real if
the average is allowed to count each pair of galaxies twice, letting the 
two shear values swap places for the second counting.  The expectation value 
of $\xi_-$ is real for shear fields that are parity invariant.  That is, if
the shear field is statistically identical after being reflected along some 
axis, then the imaginary part of $\xi_-$ has an expectation value of
0, and deviations from this value in a particular realization of a
shear field can be discarded as meaningless.

To measure the shear correlation function, we use a public, open-source software
package, called {\tt corr2}\footnote{Available at \url{https://code.google.com/p/mjarvis/}}.
It uses a ball-tree algorithm to avoid having to calculate the product of every
pair of galaxies individually.  Essentially, it calculates the shear products for groups
of galaxies that have nearly the same separation vector, and thus belong in
the same final bin.  For more details on the algorithm, see
\cite{corr2}.

Another relevant property of shear fields is that they 
can be divided into so-called $E$-mode and $B$-mode components (see \S\ref{subsec:shearfields}).  As discussed in \S\ref{subsec:sims:noise},  
in our simulated shear fields, the lensing shear is constructed to be purely $E$-mode,
whereas the shape noise (due to galaxy intrinsic shapes) is almost purely\footnote{``Almost'' because it turns out not
to be possible to make the shape noise pure $B$-mode while maintaining other
features that we wanted to have, such as a Gaussian input field and the 
galaxy shape distribution matching the real galaxy shapes.} $B$-mode.
Thus, 
separating the measurements into $E$-mode and $B$-mode components allows us
to mostly remove the largest source of noise in the measurement, which
lets us use far fewer galaxies than we would otherwise need to achieve 
a given statistical precision.

The method we use to perform the separation is called the ``aperture
mass statistic'', as discussed in \S\ref{subsubsec:metrics:variable}. 
The 
information in the shear field can be divided into $E$-mode and $B$-mode components
via the following formulae:
\begin{align}
M_E(\theta) &= \frac{1}{2} \int_0^\infty r\, \mathrm{d}r
    \left[ \xi_+(r) T_+\left(\frac{r}{\theta}\right) +
           \xi_-(r) T_-\left(\frac{r}{\theta}\right) \right] \label{eq:me}\\
M_B(\theta) &= \frac{1}{2} \int_0^\infty r\, \mathrm{d}r
    \left[ \xi_+(r) T_+\left(\frac{r}{\theta}\right) -
           \xi_-(r) T_-\left(\frac{r}{\theta}\right) \right] \label{eq:mb}
\end{align}
where
\begin{align}
T_+(x) &= \frac{x^4-16x^2+32}{128} \exp\left(-\frac{x^2}{4}\right) \\
T_-(x) &= \frac{x^4}{128} \exp\left(-\frac{x^2}{4}\right) 
\end{align}
The integrals formally go from $r=0$ to $\infty$; however, the weight functions
$T_+$ and $T_-$ go to zero quickly for large values of $x$.  At $r=0$,
$T_+$ goes to a constant, so $E$ vs. $B$ mode separation
formally requires integration over the correlation 
functions to zero separation\footnote{
See \cite{map_kilbinger} for more discussion of this difficulty for 
cosmological surveys.}.  
For GREAT3, we get around this difficulty by knowing the correlation
function of the  true input shear
field at scales smaller than the closest pairs of
galaxies in the simulation.  When we receive a submission  consisting
of the measured $\xi_+$ and $\xi_-$, we can use the
true values for the parts of these statistics that are unmeasurable from
the data.  The measured correlation functions are used for the bulk
of the range of integration, so the correction is small.

In practice, the measured correlation functions are measured only at
specific logarithmically binned values.  Thus we convert Eqs.~\ref{eq:me}
and~\ref{eq:mb} into sums over those binned values using the simplest
possible approximation (constant $\xi(r)$ within each bin).  While
this procedure would
be problematic for a cosmological analysis, leading to deviations from
the true underlying aperture masses that are more than several
percent, it is not a problem for the challenge because we can apply
the same procedure to the true input shears before comparing
with the $M_E$ and $M_B$ submitted by participants.

\section{Challenge rules}\label{sec:rules}

Here we describe the rules related to participation in the challenge.

\subsection{Teams}\label{subsec:rules:teams}

Participants can register on the leader board webpage using a user name
and e-mail address (no full name required).  The e-mail address must be a real
one, as it will be used to communicate information related
to the challenge (but will not be shared/used for any other
purposes). This is particularly important since the simulations may be
updated as needed during the challenge if problems are found, and
participants will need to know about these modifications.  At the 
close of the challenge, participants will be asked to reveal
their identities and participate in the writing of a results paper.
The first and second place winners must reveal
their identities in order to receive the prizes, and they are strongly 
encouraged to present their method at the GREAT3 final meeting, for
which travel support will be available.  During the
course of the challenge, participants are encouraged to describe their
method(s) on the wiki at the public GitHub repository described in
\S\ref{subsec:overall}.  However, before writing papers based on the
GREAT3 challenge results while the challenge is still ongoing, please
write to \url{challenge@great3challenge.info} to consult with the
leaders of the challenge.  After the challenge ends,
participants are encouraged to write papers based on the results,
preferably citing the official challenge results paper.

Submissions are to be made by teams, which can include any number of
people on them.  Likewise, people may be on any number of teams.  
Teams are permitted to submit results labeled as different methods.
On any given branch, only the top-ranked method for any team will
appear on the leader board, though the other submissions are stored
for later reference and interpretation of results.

Submissions should be considered new methods when the algorithms have
some new element involved.  A team with several methods (in
terms of algorithm or basic assumptions) may
be ranked on the leader boards for different branches with different
methods\footnote{Note that this is a change from GREAT08 and GREAT10,
  which ranked methods rather than teams.  The reason for the change
  is that, given the large variation in simulated data types (constant
and variable shear, space and ground data), we want to allow the
possibility that one team might have two or more ``specialist''
methods that only handle certain data types, but do so very well.  In
our scheme, the high rankings for those methods can be combined to
allow this team to win.}.  The points for those top rankings with different methods are
considered when determining where that team is ranked on the overall
leader board (\S\ref{subsec:leaderboard}).  
Forming a new team to submit ``new'' methods without any significant
differences\footnote{See the leader board website for 
  examples of what constitutes a different method.} in order to push other teams off the leader boards is
grounds for 
disqualification from the challenge; we reserve the right to disqualify teams for other
malicious behavior as well.  However, there is an element of choice
here: for a truly different method, people may
decide whether they wish to submit it as the same team or as a
different one.
On any given branch, teams are limited to one submission per day.  During the
course of the challenge, participants are welcome to form new teams by
opening issues on the public GitHub repository described in
\S\ref{subsec:overall}; for example, someone who only has shape
measurement code might try to form a team with someone who has PSF
estimation code in order to participate in the ``variable PSF''
branch.

Teams that include $\ge 1$ participant on the ``GREAT3 executive
committee'' (which gives them access to privileged information about
the simulations) are flagged as such.  This means that while they
appear on the individual leader boards, they do not receive an
official ranking (e.g., if their metric is at the top, they will
appear in the top position, but the first unflagged person is the one
who is ranked as first place on that board when it comes to
determining points for the overall leader board,
\S\ref{subsec:leaderboard}).  Such teams appear with starred scores on
individual branch leader boards, and do not appear
on the overall leader board at all.  The list of executive committee
members is on the FAQ at the public GitHub site\footnote{\url{https://github.com/barnabytprowe/great3-public/wiki/Frequently-Asked-Questions}}; however, it
is the responsibility of the committee members to identify themselves
as such at the time their team is formed.  If an executive committee
member wishes to join a team after the time of its formation, they
should e-mail the challenge e-mail address given in
\S\ref{subsec:overall} to change that team's status to ``flagged''.

\subsection{Overall leader board}\label{subsec:leaderboard}

To create an overall leader board, we award points to each team based on their rankings on the 
individual branches.  Each team is awarded points based on their best-ranked 5 branches (or less 
than 5, if they submit to less than 5 branches); we award 1,000 points for a fifth-place finish, 
2,000 points for fourth, 4,000 points for third, 8,000 points for second, and 16,000 points for 
first.  The team with the highest total number of points is the winner.  In the case of a tie, the total number of points from \emph{all} submitted branches will be
totaled for the tied teams, and the team with the most total points wins.  If the teams are still 
tied, then as a second-level tie-breaker they will be ranked by the
earliest submission time stamp among the branches that contributed points
to the tiebreaker (any branch in which the team placed in the top 5), with earlier time stamps 
winning over later time stamps.  We performed 10,000 Monte Carlo simulations of possible challenge 
scores, and found that our ranking method did not strongly favor specific distributions of scores, although
it rewards those who are consistent across all branches slightly more than those who specialize in 
an experiment or observation type.  Limiting to the best-ranked 5 branches helps reduce the impact 
of the number of submissions per team: around 60\% of the simulated teams submitted to 10 or fewer 
branches, and they made up around 40\% of winners, compared to 15\% of winners when we did not 
limit to the best-ranked 5 branches per team.  With this method, almost 60\% of the
first-place teams did not place first in any individual branch.

\section{Validation of GalSim shearing precision}\label{sec:shear-validation}

Here we provide evidence that GalSim can be used to create sheared
images of parametric and realistic galaxies using the DFT method of
image rendering (\S\ref{subsec:sims:rendering}) for the purpose of
testing weak lensing algorithms.

Our first test is for \sersic\ profiles.  GalSim can render \sersic\
profiles in two ways: via photon-shooting or the DFT approach.  For
the former approach, the primary approximation is the use of a lookup
table to represent the radial profile when sampling the photons.  The
shearing, convolution with a PSF, and binning into pixels is in
principle exactly represented with photon offsets.  For DFT, there are
more approximations: we have to represent the Fourier-space profile as
a lookup table, but shearing also assumes that we are in the regime
where DFTs can substitute for continuous Fourier transforms.  There
are thus additional caveats for the DFT approach, and they are in
principle independent of the issues that can arise from
photon-shooting.  Thus our first test for \sersic\ profiles is whether
sheared \sersic\ profiles agree when generated using these two
methods.  For this purpose we make images in both ways, measure their
shears using adaptive moments, and define a STEP-like calibration bias
(see \S\ref{subsec:metrics}),
\begin{equation}\label{eq:mdft}
\gamma_\text{DFT} - \gamma_\text{phot} = m_\text{DFT} \gamma_\text{phot} + c_\text{DFT}.
\end{equation}
Our target level of accuracy for shear testing with GREAT3 is that we
would like to test for calibration biases and additive systematics at
the level needed for Euclid \citep{2013MNRAS.429..661M}, $m\sim
2\times 10^{-3}$ and $c\sim 2\times 10^{-4}$.  Thus we would like our
simulation software to produce spurious shears that are a factor of
$\ge 10$ below that, i.e., $m_\text{DFT}<2\times 10^{-4}$.

\begin{figure}
\begin{center}
\includegraphics[width=0.49\textwidth]{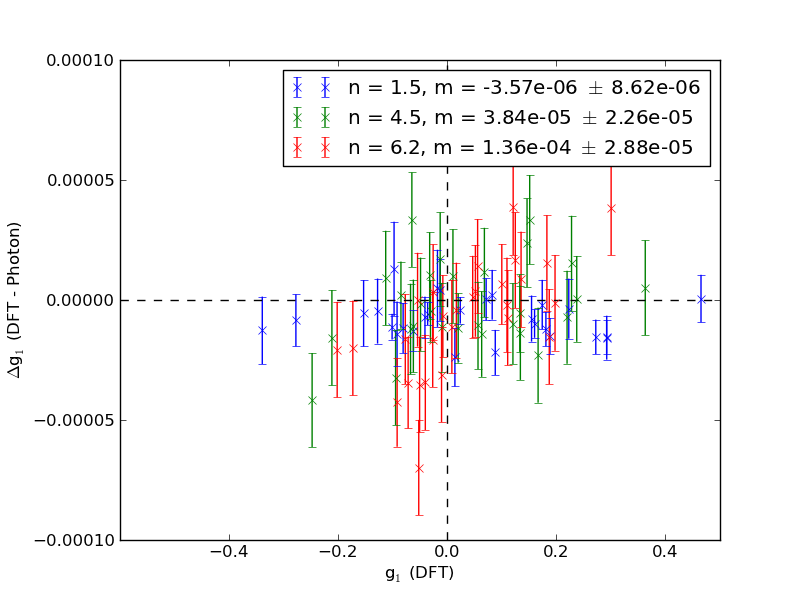}
\includegraphics[width=0.49\textwidth]{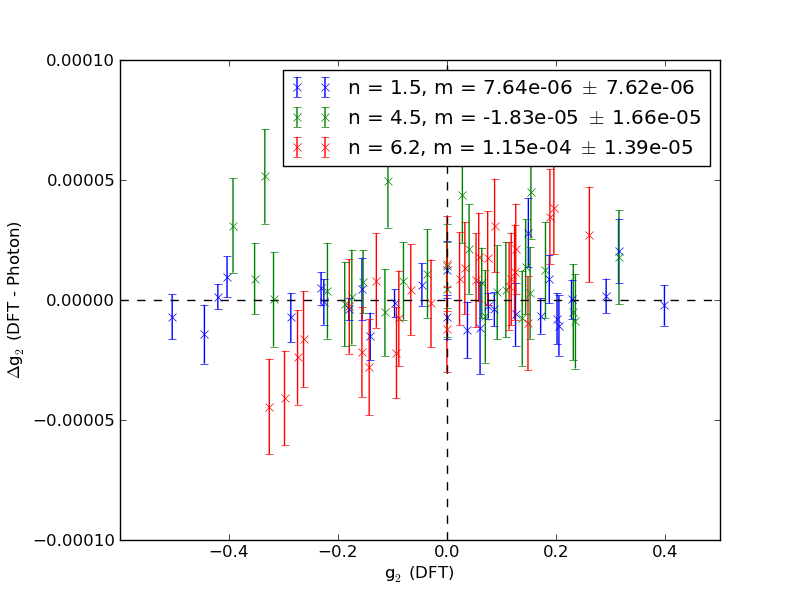}
\caption{\label{fig:sersic-validation} Difference between measured
  shears (upper panel: $g_1$; lower panel: $g_2$) for \sersic\
  profiles simulated using the two methods of image rendering in
  GalSim, photon-shooting and DFT, plotted against the shear measured
  from the DFT image. Results are shown for 30 galaxies
  with realistic size and shape distribution, and several \sersic\ $n$ 
  values shown in the legend.  The points can be fit to lines to
  measure $m_\text{DFT}$ as defined in
  Eq.~\ref{eq:mdft}, and the best-fit values are shown in the legend.}
\end{center}
\end{figure}

We carried out a test of shearing accuracy for each of the two shear
components, alongside a similar test for the correct rendering of
galaxy size.  Several values of \sersic\ indices $n$ were
investigated, for a range of galaxy half light radii and intrinsic
ellipticities $\varepsilon^{(s)}$ drawn from a random sample of 30
single component S\'{e}rsic model fits to the COSMOS training data
sample.  For each galaxy a circular profile was first sheared to
create an object with ellipticity $\varepsilon^{(s)}$, convolved with
a COSMOS-like PSF, then rendered as an image via both DFT and photon
shooting.  Differences between moment estimates of the resulting
ellipticities are plotted in Fig.~\ref{fig:sersic-validation}.  As
shown, the values of $m_\text{DFT}$ demonstrate that we can
consistently represent galaxy shears at the few $\times 10^{-6}$ level
for $n=1.5$, with $m_\text{DFT}$ rising as high as $\sim 1\times
10^{-4}$ for the highest $n=6.2$ (note that for GREAT3 we use $n\le
6$).  These values are safely below our target values of $2\times
10^{-4}$.  Since errors in DFT and photon-shooting are completely
independent, it is highly improbable that this good agreement is due
to chance, and it supports our claim that we can accurately shear
galaxies rendered via DFT for the GREAT3 challenge.

As a parallel investigation, we also estimated the size of
the galaxies in the images described above, using adaptive
moments.  By fitting a slope to the differences between DFT and photon
shooting results, we can also estimate the accuracy at which weak
lensing magnifications can be simulated using GalSim.  We found a
slope of $m_\text{DFT} = (4.7 \pm 0.7) \times 10^{-5}$ for the
\sersic\ $n=1.5$ galaxy sample, $m_\text{DFT} = (-1.4 \pm 3.6) \times
10^{-5}$ for the $n=4.5$ sample, and $m_\text{DFT} = (-3.4 \pm 8.1)
\times 10^{-5}$ for the $n=6.2$ sample.  These results are safely
below the $2 \times 10^{-4}$ target adopted for multiplicative-style
biases in the simulation of shear.  Indeed, the signal-to-noise
expected for cosmological magnification measurements has been
estimated as $\lesssim 50 \%$ relative to shear (e.g.\
\citealp{waerbeke10,2012ApJ...744L..22S,2013arXiv1306.6870D})
motivating a corresponding relaxation of requirements by a factor
$\gtrsim 2$ (although this figure is dependent both on the dataset and
analysis technique used, see e.g.\ \citealp{2011arXiv1111.1070H}).
These results suggest that the representation of galaxy sizes with
GalSim therefore falls comfortably within requirements for future
surveys.

Next, we show that we can accurately shear more complex,
realistic galaxy images.  For the GREAT3 challenge, we must remove the
{\em HST} PSF, shear and magnify, and convolve with the target PSF.  In this
case there is no ground truth. Instead, we begin with a simpler test for
which we do have ground truth: we treat the {\em HST} PSF for our training
galaxies as part of the galaxy itself.  In that case, we can compare
the shape of the original image ({\em HST} PSF included) with the
shape when we shear it by a small, known amount.  We carry out this
test for simulated \sersic\ profile images and for realistic {\em HST}
galaxies, and ensure that the recovered shears are as expected despite
the need to carry out interpolation to do the shearing.  In this case, since the original galaxies are not round, we
calculate the difference between the observed shear (after applying a
shear) and the expected one given the intrinsic shear and the applied
one, and define calibration bias due to interpolation $m_\text{interp}$ as
\begin{equation}\label{eq:mint}
\gamma_\text{obs}-\gamma_\text{expected} = m_\text{interp} \gamma_\text{applied}
+ c_\text{interp}.
\end{equation}
We find for $m_\text{interp}$ for the two shear components is
$(-1.3\pm 0.5)\times 10^{-5}$ and $(-0.2\pm 3)\times 10^{-5}$.  These are both safely below our
tolerance for spurious shear in the simulation process.
These numbers come from using the default interpolants and settings in
GalSim; future work might involve refining these, but they are clearly
adequate for the levels of calibration bias that can be detected in
GREAT3.  $c_\text{interp}$ is of order $10^{-10}$, consistent
with zero within the errors.

In addition,
we check for leakage between shear components (i.e., that applying
one shear component does not result in an incorrect level of shear in the other
component).  For example, we define cross terms such as
\begin{equation}
\gamma_\text{obs,1} - \gamma_\text{expected,1} = m_\text{interp}^{(1,2)}\gamma_\text{applied,2}
\end{equation}
and likewise for leakage between magnification and shear.  We find
that $m_\text{interp}^{(1,2)}$ and $m_\text{interp}^{(2,1)}$ are of
order $1\times 10^{-5}$.

Our final test is to show that we can successfully carry out the
process of ``reconvolution'' \citep{2012MNRAS.420.1518M} using
GalSim - that is, when we say we are representing what some galaxy
looks like with an added shear $\gamma_\text{applied}$ when viewed at lower resolution, is
that statement correct?  This test was carried out using
simulated \sersic\ profiles at high resolution, putting them through
the reconvolution process and ensuring that the result looks like what
we simulate by taking the original \sersic\ profile and viewing it
directly at low resolution.  We quantify any error in the effectively
applied shear due to the reconvolution process as $m_\text{reconv}$,
defined by
\begin{equation}
\gamma_\text{reconv} - \gamma_\text{direct} =
(1+m_\text{reconv})\gamma_\text{applied}. 
\end{equation}
$m_\text{reconv}$ was determined for 270 galaxies randomly selected
from the training
sample described in Appendix~\ref{sec:galaxies}, for a space-based and
a ground-based target PSF.  As for previous tests, our target value is
$m_\text{reconv} < 2\times 10^{-4}$.  Since galaxies with different light
profiles might be more or less difficult to accurately render using
reconvolution, we consider not only the mean $\langle m_\text{reconv}\rangle$
but also its standard deviation, as an indicator of possible galaxy
types for which the method fails to work sufficiently accurately even
if it works for most galaxies.  For the default GalSim settings used
for the GREAT3 simulations, we find $\langle
m_\text{reconv}\rangle$ is completely consistent with zero, with a
standard deviation of $3\times 10^{-5}$, well below our target value
of $m_\text{reconv}$ of $2\times 10^{-4}$.  This result shows that any
profile inaccuracies due to the reconvolution process do not interfere
with our ability to accurately render what a galaxy looks like with a
particular shear, even for different galaxy types.

The results in this section use the default set of parameters for DFT
and photon-shooting accuracy in GalSim; more detailed investigations
will be presented in \galsimpaper.

\section{Real galaxy dataset}\label{sec:galaxies}

Here we describe the dataset used to simulate a realistic galaxy
population in the GREAT3 challenge.

\subsection{{\em HST} training sample}

The training sample that is compiled here comes from the COSMOS
survey, using galaxy selection criteria from \cite{2012MNRAS.420.1518M},
as summarized below.

The COSMOS {\it Hubble Space Telescope} ({\it HST}) Advanced Camera
for Surveys (ACS) field
\citep{2007ApJS..172..196K,2007ApJS..172....1S,2007ApJS..172...38S} is
a contiguous 1.64 degrees$^2$ region centered at R.A.$=$10:00:28.6,
Dec.$=$+02:12:21.0 (J2000).  Between October 2003 and June 2005 ({\it HST}
cycles 12 and 13), the region was completely tiled by 575 adjacent and
slightly overlapping pointings of the ACS Wide Field Channel. Images
were taken through the wide F814W filter (``Broad I''). We use the
`unrotated' images (as opposed to North up) to avoid rotating the
original frame of the PSF. The raw images are corrected for charge
transfer inefficiency (CTI) following
\citet{2010MNRAS.401..371M}. Image registration, geometric distortion,
sky subtraction, cosmic ray rejection and the final combination of the
dithered images are performed by the multidrizzle algorithm
\citep{2002hstc.conf..337K}. As described in
\citet{2007ApJS..172..203R}, the multidrizzle parameters have been
chosen for precise galaxy shape measurement in the co-added images. In
particular, a finer pixel scale of $0.03\arcsec/$pix was used for the
final co-added images ($7000\times 7000$ pixels).

The following cuts are then applied on catalogs derived from the
COSMOS images; for more details on the flags, see \cite{2007ApJS..172..219L}:
\begin{itemize}
\item $F814W< 25.2$: This cut corresponds to a $S/N$ limit of $\sim
  20$.  However, as discussed in \S\ref{subsec:sims:deep}, we only
  use those galaxies at $<23.5$ for GREAT3, applying simple
  transformations (\S\ref{subsec:fake}) to mimic a fainter sample.
\item MU\_CLASS $=1$: This requirement uses the relationship between
  the object magnitude and peak surface brightness to select galaxies,
  and to reject other objects.
\item CLEAN $=1$: This cut is required to eliminate galaxies with
  defects due to very nearby bright stars, or other similar issues.
\item GOOD\_ZPHOT\_SOURCE $=1$: This cut requires that there be a good
  photometric redshift, which typically is equivalent to requiring
  that the galaxy not be located within the masked regions of the
  ground-based $BVIz$ imaging used for photometric redshifts.  We
  impose this cut here because we wish to test the galaxy population
  going into our simulations to ensure that it is representative of
  reality, and having a photometric redshift estimate is an important
  part of those tests.
\end{itemize}

Following the procedure in \cite{2012MNRAS.420.1518M} for a brighter
subset of the data, postage stamps were cut out around the position of
each galaxy.  The background level was subtracted, and additional
objects besides the central one were masked with a correlated noise
field with the same properties as the noise in the rest of the image.
As in that work, in order to remove the effects of the COSMOS PSF, we
use PSF models from a modification of version 6.3 of the Tiny Tim
ray-tracing
program\footnote{\url{http://www.stsci.edu/software/tinytim/}}.
These models represent PSFs for different primary/secondary
separation, since that separation is the main determinant of the PSF
ellipticity; while imperfect particularly at long wavelengths
\citep{1998SPIE.3355..608S}, the Tiny Tim PSFs are close enough to
reality to use in our simulations.  Future work will include
empirically-estimated PSFs.

As described in e.g.~\cite{2008MNRAS.386..781M}, the COSMOS field is
small enough that, when measuring quantities as a function of redshift
in small redshift bins,
large-scale structure in the field induces non-negligible noise in the
results.  This should also be the case when using it 
as a training sample to estimate shear calibration as a function of
redshift: the intrinsic ellipticity distribution can differ in dense and
underdense environments due to their different galaxy populations, so for narrow redshift slices, the shear calibration
would reflect those different populations.  However, here we are using
the COSMOS sample to measure the shear calibration for some
redshift-averaged population, such that the large-scale structure
fluctuations in narrow $\Delta z$ slices effectively cancel out.  As a
result, we do not impose any density-dependent weighting on the sample.

\subsection{Parametric fits}\label{subsec:fits}

We fit the galaxies in the training set with parametric models. The functional form is given by a \sersic{} profile \citep{Sersic69}. The radial surface brightness profile is 
\begin{eqnarray}
I(R) &=& I_{1/2}\exp\left[-b_n\left(\left(R/\Reff\right)^{1/n}-1\right)\right]\quad ,\\
R &=&\left[\left(\left(x-x_0\right)\cos\phi+\left(y-y_0\right)\sin\phi\right)^2\right.\\	
&+&\left.\left(\left(y-y_0\right)\cos\phi-\left(x-x_0\right)\sin\phi\right)^2/q^2\right]^{1/2}\quad,\nonumber
\end{eqnarray}
where \Reff{} is the half-light radius, $I_{1/2}$ is the surface
brightness at the half-light radius, $n$ is the \sersic{} index, and
$b_n$ is a normalization factor dependent on the \sersic{} index. The
radius, $R$, defines an ellipse, with minor-to-major axis ratio $q = b/a$.  We fit each galaxy twice: once with a \sersic{} profile, and once with a de Vaucouleurs ($n=4$) bulge profile plus an exponential ($n=1$) disk profile. The fitting method is described in detail in \cite{2012MNRAS.421.2277L}. The \sersic{} profile contains 7 free parameters: \Reff{}, $I_{1/2}$, $n$, the central position (2 parameters), the axis ratio of 
elliptical isophotes, and the position angle. We place some
constraints on the fitted parameters. The surface brightness must be
positive, the \sersic\ index is between 0.1 and 6 (following
\citealt{Blanton2005a}), the axis ratio $0.05\le q\le 1$, and the size of the galaxy must be smaller than
the the size of the postage stamp. As in \cite{2012MNRAS.421.2277L}, the
\sersic\ models cutoff smoothly at large radii. The cutoff radius varies
smoothly from 4 half-light radii for $n=1$ to 8 half-light radii for
$n=4$.

The bulge$+$disk models have 10 free parameters, since we fix the
\sersic\ indices of both components ($n=4$ for the bulge and $n=1$ for the
disk) and require that the bulge and disk share the same centroid. In
addition, we require that the bulge half-light radius is less than
that of the disk. Previous
studies have shown that varying the bulge \sersic\ index does not yield
statistically significantly better fits for the typical galaxy in this
sample \citep{Simard2011, 2012MNRAS.421.2277L, deJong1996}.

The best-fit parameters are found using a 2-dimensional
Levenberg-Marquardt minimization, \verb+mpfit2dfun+ in IDL
\citep{Markwardt09}. The fitter minimizes the weighted sum of the
squared differences between the galaxy image and PSF-convolved
model. This $\chi^2$-minimization method assumes the pixel values are
uncorrelated, which is not true for the {\em HST} postage stamps.  The
weights are given by the inverse variance in each pixel, including sky
noise and photon noise from the source.  Although this is the optimal
weighting scheme for a least $\chi^2$ fit, it does introduce changes
in the weighting scheme as functions of galaxy brightness. Faint
galaxies are fit with constant, sky-noise-dominated weights, while
bright galaxies are down-weighted in the central regions. The initial
values for the minimization are obtained from an exponential profile
fit to the galaxy. 

We have tested the fits by creating mock
single-\sersic\ and bulge+disk images at various resolutions and
$S/N$ using GalSim. The fitter recovers the correct
input parameters for the relevant range of resolution and
$S/N$, although the uncertainties grow as the
$S/N$ decreases (Lackner, et al. in prep). For simulated galaxies where the ratio of bulge flux to total flux ($B/T$) is between $0.3$ and $0.7$, the error in $B/T$ varies from $0.10$ to $0.17$ as the $S/N$ decreases from $100$ to $50$, typical for galaxies in the HST training sample. The errors in bulge half-light radius are typically $25-44\%$, while the errors in the disk half-light radius are always smaller, $10-30\%$. These uncertainties depend most strongly on $S/N$, not $B/T$, when $B/T$ is far from both $0$ and $1$. For single-component simulated galaxies, the uncertainties in galaxy sizes are smaller still, ranging from $2-10\%$, depending on $S/N$, input galaxy size, and S\'ersic index. The errors in size are usually $3$ times larger for galaxies with S\'ersic index near $4$ than for bulge-less exponential disks. In all cases, down to $S/N=50$, the mean offsets between the measured values and the input parameters are within $1$ standard deviation. 

For our simulations, we use the bulge$+$disk model described above,
except for cases where $B/T$ is below
$0.1$ or above $0.9$; or where the bulge radius or axis ratio runs up against the fit limits (e.g., $q_{\rm bulge}=0.05$ precisely).  In those cases, the galaxy is dominated by a
single component, and it occasionally happens that the subdominant
component has extremely large radius and low surface brightness
(absorbing some sky gradient), or otherwise poorly constrained
parameters.  Thus we do not use the bulge$+$disk
fits for these edge cases, and instead use the single \sersic\
fits with free $n$.  We also require that the median absolute
deviation or MAD be lower for the $2$-component fits, otherwise we
just use the single component fits.  After all cuts, we use
two-component fits for the $\sim 1/3$ of the sample for which they
seem justified, and single \sersic\ fits for the rest. Note that the fits do include
populations of galaxies with $0.1<B/T<0.2$ with bulges with unusual
properties (e.g., $q\sim 0.1$); visual inspection suggests that these
are not fit failures but rather the fitter attempting to represent
bars or the beginnings of spiral arms using a ``bulge'' component.
Thus we do not attempt to remove these fits.

\begin{figure}
\begin{center}
\includegraphics[width=0.34\columnwidth,angle=0]{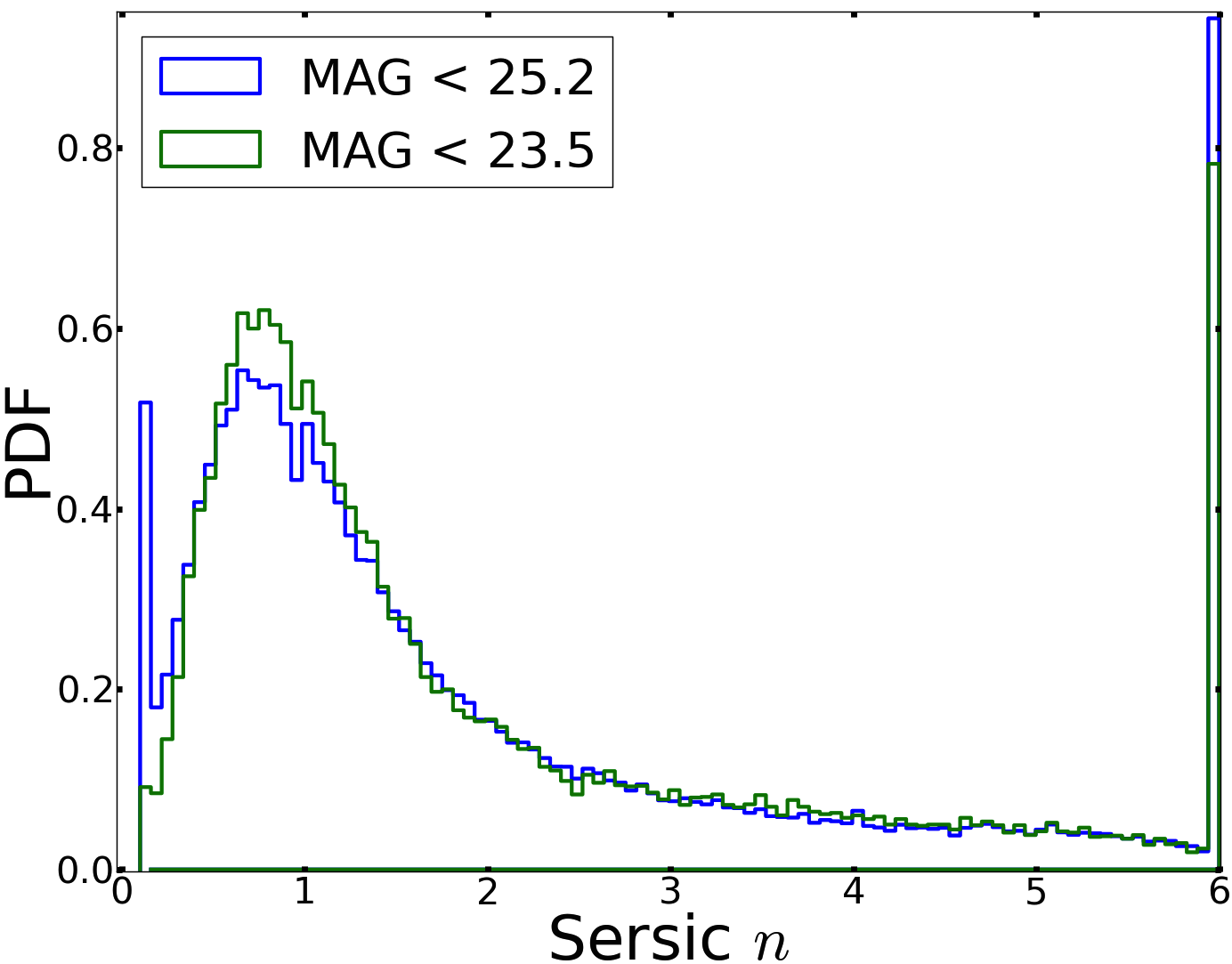}
\includegraphics[width=0.34\columnwidth,angle=0]{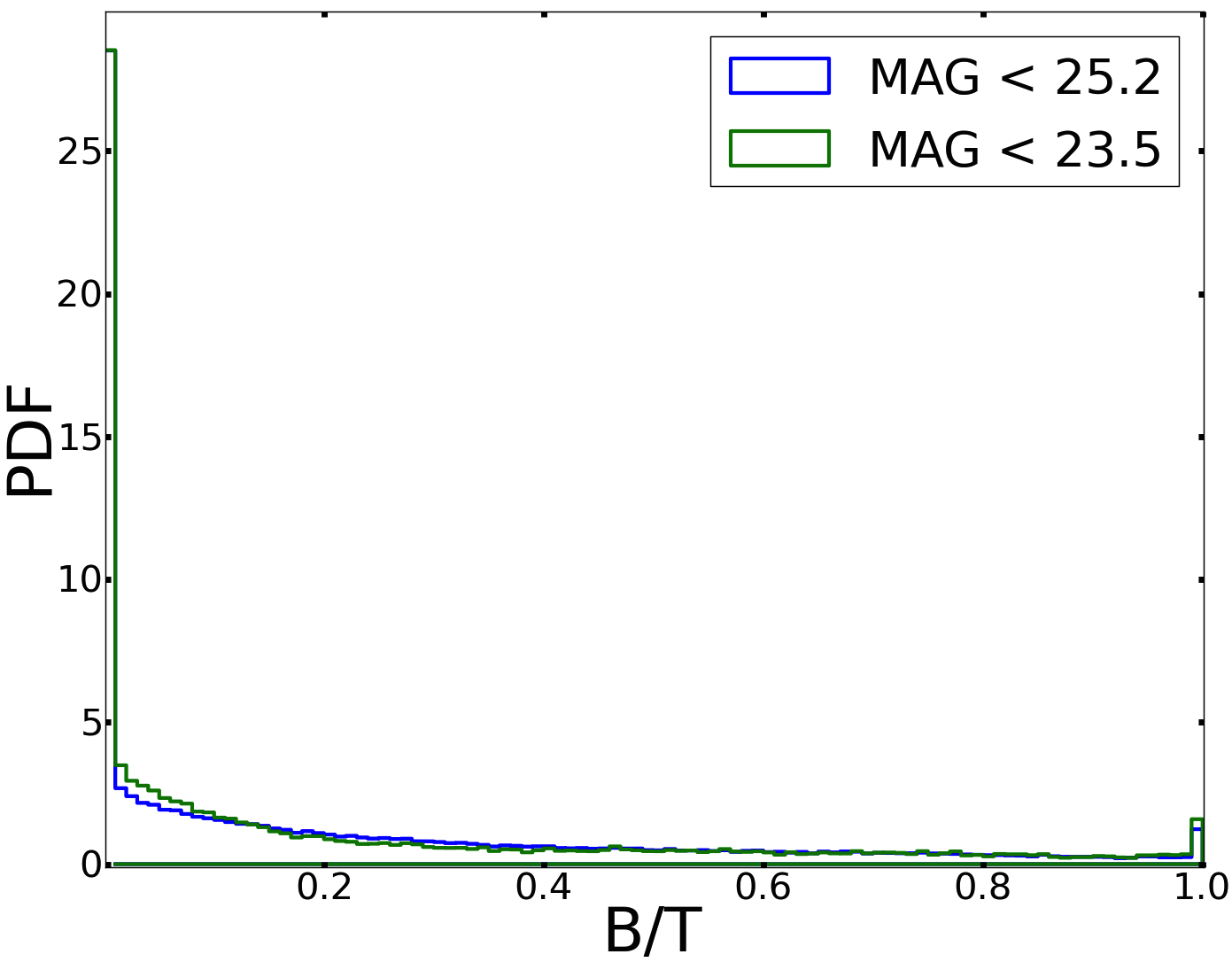}
\includegraphics[width=0.34\columnwidth,angle=0]{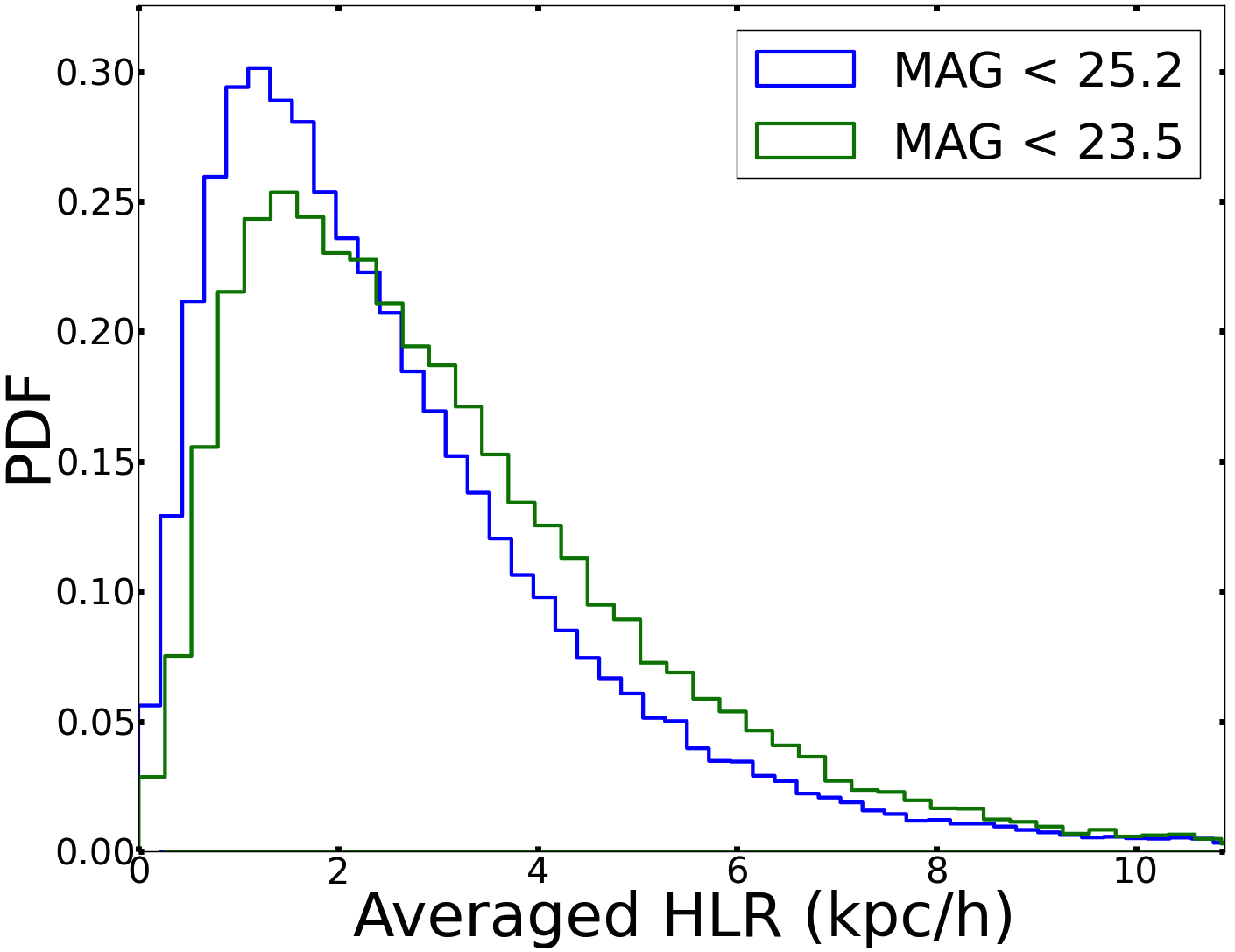}
\includegraphics[width=0.34\columnwidth,angle=0]{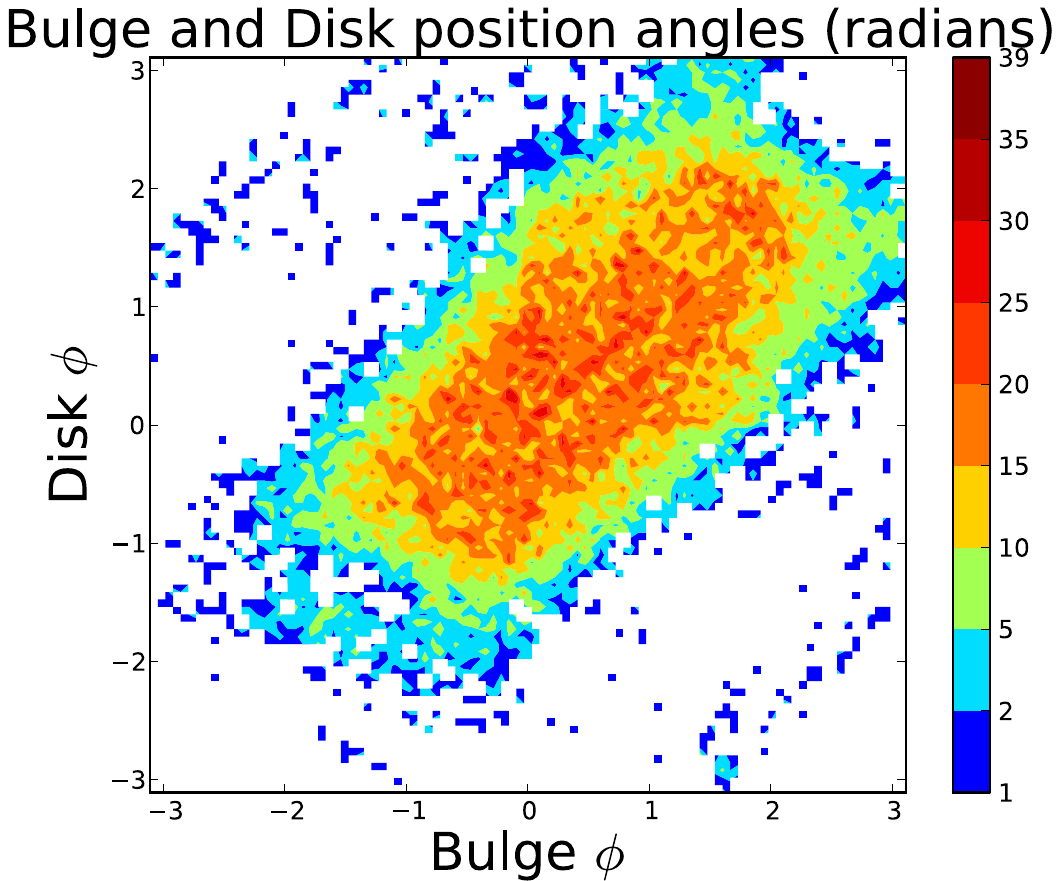}
\includegraphics[width=0.34\columnwidth,angle=0]{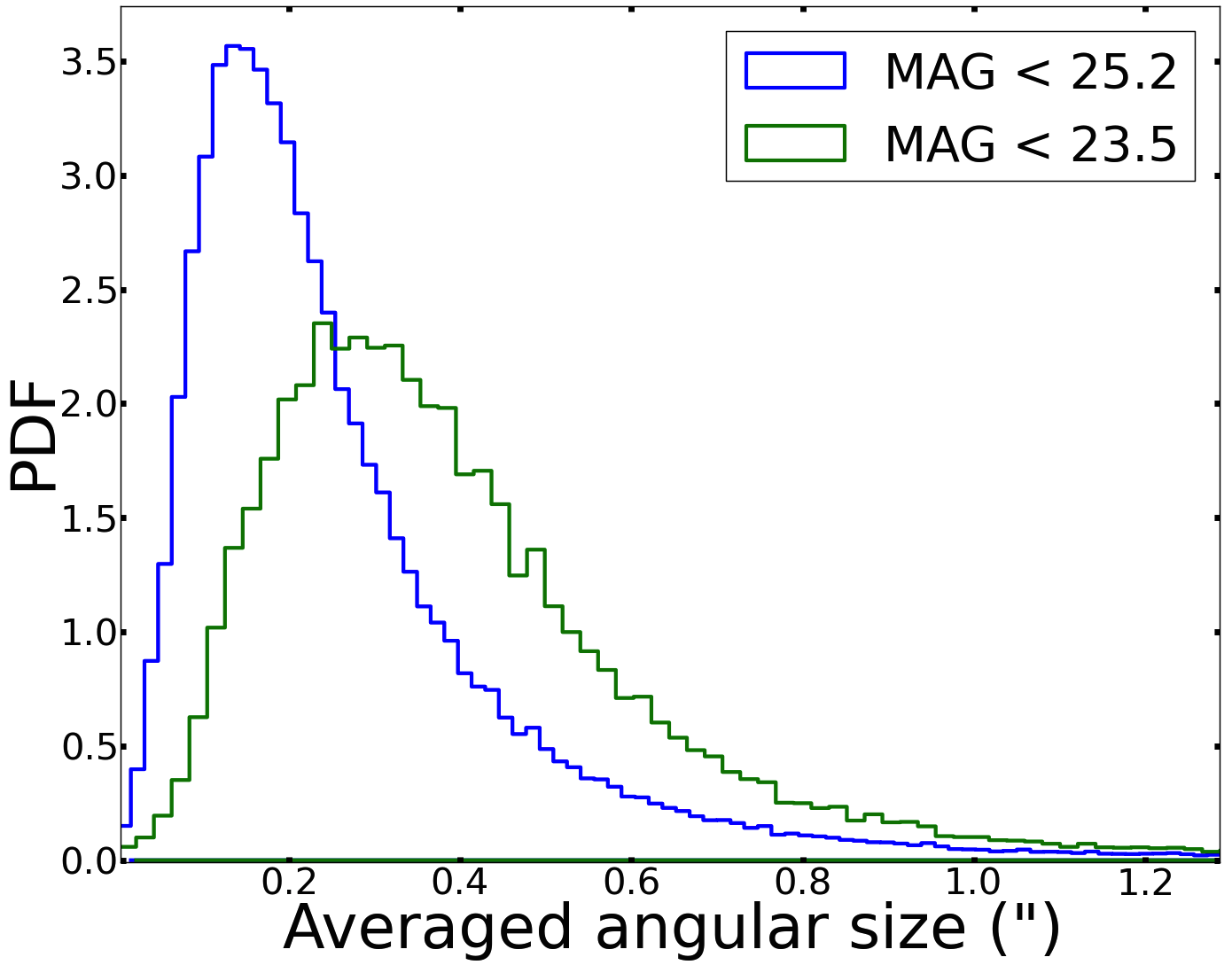}
\caption{\label{fig:fits}For two magnitude cuts, we show distributions
of various galaxy properties from the fits described in
appendix~\ref{subsec:fits}. \textbf{Top left:} Distribution of
\sersic\ $n$ values, with a slight tendency to pile up at the lower
and upper limits for the fainter sample (due to noise).  \textbf{Top
  right:} Distribution of bulge-to-total flux ratio $B/T$ from the
two-component fits.  \textbf{Middle left:} Distribution of physical
half-light radius, which tends to smaller values for the fainter
sample because that one has more low luminosity
objects. \textbf{Middle right:} For the sample limited at
25th magnitude, for galaxies with two significant components, this
plot shows density contours for the joint 
distribution of bulge and disk position
angles. \text{Bottom:}Distribution of angular half-light radius.}
\end{center}
\end{figure}

Fig.~\ref{fig:fits} shows some properties of the sample based on these fits.  Of particular note is the middle right 
panel, which compares the position angles for bulge and disk shapes in
the two-component fits.  As shown, the bulge and disk have a
significant tendency to be aligned with each other, but some
non-negligible offsets are allowed and thus will be represented in our
simulations even for the control experiment.

\subsection{Mimicking a fainter sample}\label{subsec:fake}

Since we wish to use a sample with robust two-component fits to make
the simulations, but want to simulate a galaxy sample that is typical
for deeper surveys ($F814W\sim 25$), we have used a sample with
$F814W<23.5$ to mimic a deeper sample with a limit of $<25.2$.  In
details of intrinsic properties like redshift, this would be a
difficult task.  However, we largely wish to reproduce the {\em
  observed} properties of the sample that determine shear calibration,
including the distributions of $S/N$, apparent size, intrinsic
ellipticity, and morphology.  To test our ability to do this, we use
the fewer-parameter single-\sersic\ fits (and the $B/T$ from the
double \sersic\ fits described in the above subsection) and
demonstrate that a very simple prescription enables us to achieve our
goal.

\begin{figure}
\begin{center}
\includegraphics[width=0.6\columnwidth,angle=0]{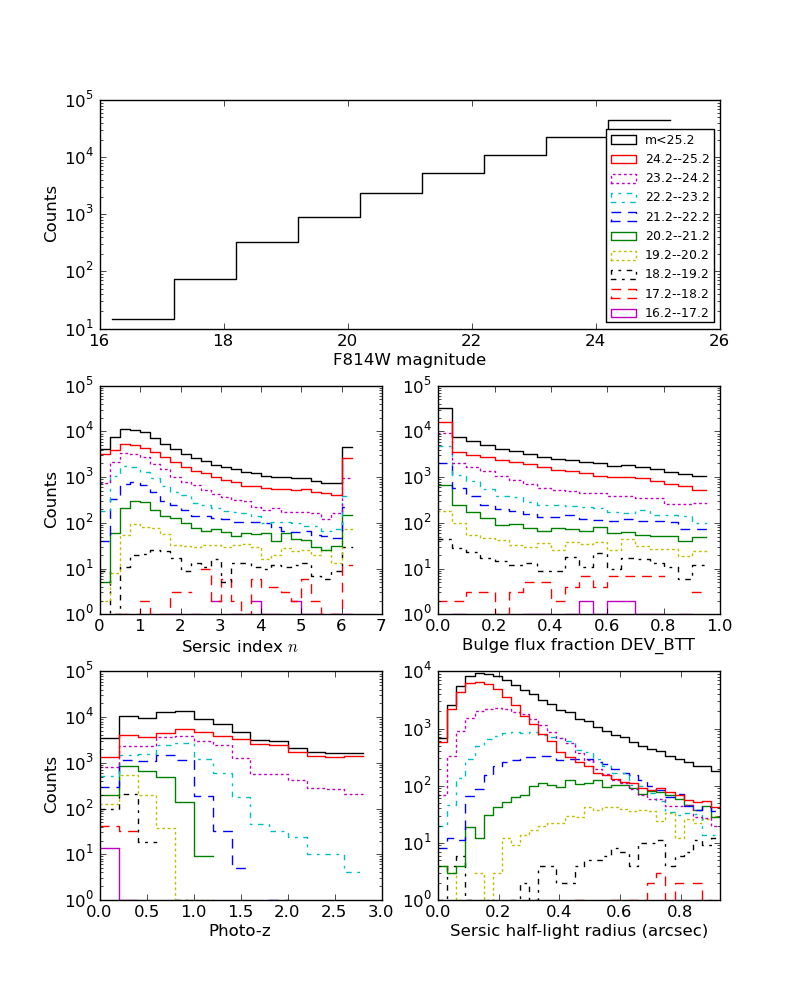}
\caption{\label{fig:properties-vs-mag}Histograms of the properties of
  training sample galaxies for magnitude bins defined in the uppermost
  panel.}
\end{center}
\end{figure}

Fig.~\ref{fig:properties-vs-mag} shows the distributions of $B/T$,
\sersic\ $n$ and half-light radius, and photo-$z$ from the
catalogs of \cite{2012ApJ...744..159L}. The histograms of these
properties are shown as a function of magnitude in bins shown in the
upper left panel, and for the sample overall in the solid black line.
As shown, the histograms of $B/T$, \sersic\ and $n$ are
largely independent of magnitude\footnote{Another important property, the
ellipticity distribution, is also consistent with being independent of
apparent magnitude (modulo noise, which increases scatter towards high
ellipticity in a well-understood way; e.g.,
\citealt{2007ApJS..172..219L}). We have not plotted this quantity,
because the ellipticity distribution is so central to shear inference,
and as for real data, challenge participants must carry this out
themselves from the simulated data.}.  However, we see the expected trends
that fainter galaxies are (a) at higher redshift and (b) smaller in
size.  As noted previously, (a) does not affect shear
calibration per se, but rather the true shear experienced by a galaxy;
hence if we are trying to calibrate some average shear calibration, we
do not need to reproduce distributions of photo-$z$.  Therefore, our key
challenge is to get the size and $S/N$ distribution of the $<23.5$ sample to
look like that of the $<25.2$ sample.

\begin{figure}
\begin{center}
\includegraphics[width=0.8\columnwidth,angle=0]{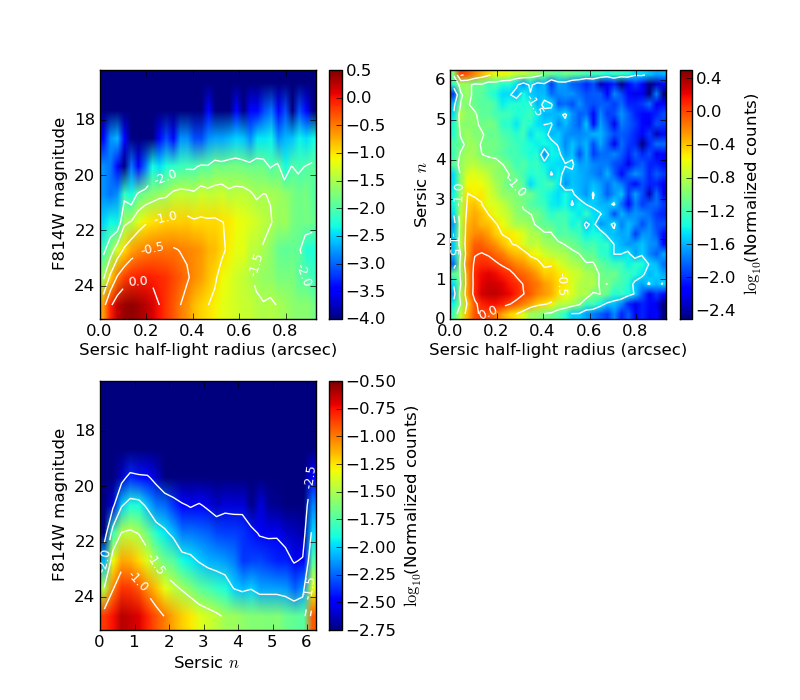}
\caption{\label{fig:2d-properties}Two-dimensional contour plots
  showing the relationship between $F814W<25.2$ training galaxy properties, shown
  on a logarithmic scale.}
\end{center}
\end{figure}

We find that a simple mapping that involves reducing the flux and
decreasing sizes by a factor of $0.6$ is sufficient to
make the $<23.5$ sample look statistically like the $<25.2$ sample in
terms of the distributions of apparent size, $S/N$, $n$, $B/T$, and
$\varepsilon$.  The 2d distributions of properties for the $<25.2$
sample (Fig.~\ref{fig:2d-properties}) are almost completely
reproduced by the $<23.5$ sample if we make this transformation.  The
one exception to this statement is a slight difference in the
ellipticity distribution; the ``fake'' sample has fewer
high-ellipticity objects.  However, 
since the existence of those high-ellipticity objects in
the faint sample is consistent with being caused by noise, it may actually be 
a benefit that our ``fake'' deep sample does not contain them.

Thus, for all branches, we always
apply this transformation factor of $0.6$ to the observed sizes in the
$I<23.5$ sample to
mimic a deeper sample.

We have not confirmed that this scheme reproduces the fraction of irregular
galaxies, since we have no good way of quantifying irregularity for
the faint sample.  It is likely that our
procedure slightly under-represents the population of irregulars,
which is known to increase at higher redshift.  This means that our
conclusions about realistic galaxy morphology might slightly
underestimate that in reality.  However, this seems preferable to the
alternative of using rather noisy galaxy images as the basis for our
simulations, effectively considering the non-negligible noise as part
of the galaxies.
  

\section{Optical PSF models}\label{sec:optpsf}

Here we describe the specific optical PSF models used for the
``variable PSF'' experiment, for simulated ground- and space-based
data.

For the simulations that mimic observations from a space telescope, we
have secured an approximate description of the design residual (\S\ref{subsubsec:sims:psf:opt}) of a
prototype telescope model for the 2.4m WFIRST-AFTA
mission\footnote{\url{http://wfirst.gsfc.nasa.gov/}} (\citealt{content-wfirst}).  The model consists of a Zernike polynomial
description of wavefront errors up to order $j=11$ in the notation of
\citet{1976JOSA...66..207N}, and therefore contains trefoil and
third-order spherical aberration (but no higher order aberrations). 
This Zernike approximation to the design residual was
provided at a series of fixed locations in the
WFIRST-AFTA FOV, and we interpolate the Zernike terms between these
locations to provide a fully continuous approximate model of a space
telescope.   Additional aberrations, to model those due to
misalignment or figure errors, were included as additions to these Zernike
terms (also up to order $j=11$ only).  Values of these
additional Zernike aberrations were chosen such that the ensemble
root mean square wavefront error added was $\lambda / 13$, where
$\lambda$ is the wavelength of the light being observed.  This is a
relatively stringent operational definition of a  diffraction-limited
optical system, and a target for aberrations due to misalignment and
figure errors for space missions such as WFIRST-AFTA.

In addition to these aberrations, the WFIRST-AFTA prototype model includes 6
non-radial struts (i.e., ones that do not go directly across the center of
the aperture).  GalSim is currently only able to simulate PSFs with
radial struts, so that is the
model we use for GREAT3.  Since jitter can be directional but
typically not with a preferred direction over long time-scales, we
model jitter as convolution with a Gaussian with RMS of
$0.005$--$0.015$\arcsec\ per axis, with ellipticity from 0 to 0.3 but
random direction.  In contrast, charge diffusion often has some
preferred direction, so we model it as a Gaussian with
$\sigma=0.05$-$0.2$ pixels, with ellipticity from $0$-$0.2$, always in
the same direction.   Like the additional aberrations, the jitter and
charge diffusion parameters are chosen for each field as a whole, and
for a given epoch, they are the same for all subfields within the field.

We need to determine a size for the tiles within the $10\times 10$
deg$^2$ images that will represent individual fields of view for the
PSF model.  The WFIRST-AFTA model is defined within a $0.42\times 0.42$
deg$^2$ FOV; we artificially stretch these length scales to $0.5\times
0.5$ deg$^2$, which means we can tile a $10\times 10$ deg$^2$ region
with $400$ PSF tiles in the space-based simulations.

For the simulations that mimic observations from a ground-based
telescope, we use an approximate description of the design residual of
an early model\footnote{S. Kent and M. Gladders, priv. comm.} for the Dark Energy Camera (DECam) at the Blanco
Telescope in Chile.  This model differs in some
respects from the one that was actually used, but it is nonetheless a
reasonable optical PSF model for an instrument on a 4m telescope.  As
for the WFIRST-AFTA PSF model, we restrict ourselves to a Zernike
polynomial description of wavefront errors up to order $j=11$ at a
series of fixed locations in the FOV, between which we interpolate the
Zernikes.  We add additional
aberrations to the ground-based PSF model to represent misalignment and tilt, based on a
model for DECam determined using extra-focal imaging (code to be
included in a future version of GalSim).  This model is defined over a
$1.56\times 1.56$ deg$^2$ field of view; however, for convenience, we
stretch all length scales so that it is $2\times 2$ deg$^2$, which
allows us to use 25 optical PSF tiles within a $10\times 10$ deg$^2$
image.

\section{Design and implementation of the atmospheric PSF model}  \label{sec:atm_psf}

\subsection{The PhoSim atmospheric model}

Here we give more details on the atmospheric model used for PhoSim,
which we use as the basis for GREAT3 as described in
\S\ref{subsubsec:sims:psf:atm}. In this model, a set of frozen Kolmogorov screens 
\citep{Kolmogorov1941} are distributed vertically above the telescope 
(representing the column of air above the telescope). 
For the work described here, the atmospheric
model assumes 7 atmospheric layers at altitudes of 16km, 8km,
4km, 2km, 1km, 0.5km, and 0.02km (ground layer), each having different strengths.
During the time 
of the exposure, the screens move according to the wind conditions 
at different altitudes. As photons propagate through different parts of the 
screen at different times, their trajectories are perturbed by an 
amount depending on the wavelength and the value of the screen 
at that location. This simulates the refraction of light 
as it passes though air of different densities (and thus refractive 
index). Atmospheric dispersion is included by scaling this 
perturbation according to the wavelength and zenith angle, as the 
screens represent a thicker layer of air when the telescope is pointed 
away from zenith.

The ``frozen screen approximation'' is justified since the time scale 
for the shapes of turbulent cells to change significantly is much longer 
than the time required for those cells to pass through the field 
of view, given the typical wind speeds of a few meters per second 
\citep{1938RSPSA.164..476T, 2009JOSAA..26..833P}. These 
atmospheric screens are constructed according to a full 
three-dimensional van Karman power spectrum 
\citep[see, e.g.,][]{1994ewpt.book.....S} 
with assigned parameters including the structure function, inner scale, 
outer scale, wind speed and wind direction. Adopting the model of 
\citet{2006MNRAS.365.1235T}, PhoSim uses 7 atmospheric layers 
(including ground layer), each layer has the effective physical size 
of $\sim2.6 \times 2.6$ km$^{2}$ and resolution of $\sim 1\times 1$ 
cm$^{2}$. Since storing all the information in these large turbulent 
screens while ray-tracing is practically impossible, PhoSim adopts 
the technique used by \citet{2008WRCM...18...91V} and splits the full 
van Karman power spectrum into three, each containing a smaller 
range of scales. PhoSim then generates three much smaller screens 
with these piecewise power spectra and only registers the value of the 
``combined screen'' on the fly as the photon hits a specific pixel on the 
screen. 

Since the specific model parameters vary from site 
to site, PhoSim uses parameters based on 
atmospheric data taken close to the LSST site, Cerro Pachon, Chile 
(2738 m above sea level, $70^{\circ}44'01''$W, $30^{\circ}14'17''$S)
in order to simulate LSST data.  
Since this site also hosts the 8-meter Gemini-South telescope and the 
4-meter SOAR telescope, which are both equipped with adaptive optics 
instruments, relatively complete atmospheric data and literature can be 
found \citep{Vernin1998, 2004A&A...416.1193A, 2000SPIE.4007.1088E}. 
Note, however, that dynamic information about the variation in timescales 
shorter than a day is currently lacking. 

\subsection{Estimation of atmospheric PSF parameters}

In this section we explain how we derived the ellipticity values and 
the spatial variation of the ellipticity and size of the atmospheric PSF 
using PhoSim.

We used PhoSim to make simulated images of exposure 
time\footnote{Note that real observations are typically anywhere from 
one to several minutes long, though PhoSim typically runs with 15s 
exposures to match the LSST observation plan.} 10s, 20s, 60s, 120s, 
each covering $0.5\times 0.5$ deg$^2$, 
with stars that have $S/N\sim1000$
on a regular grid of $0.5$ arcmin grid spacing.  For each exposure time
we made 10 $r-$band images with different seeing/wind/structure
functions based on a random seed. 
We then carried out several tests on the resulting images.

\begin{figure}
\begin{center}
\includegraphics[width=0.5\columnwidth,angle=0]{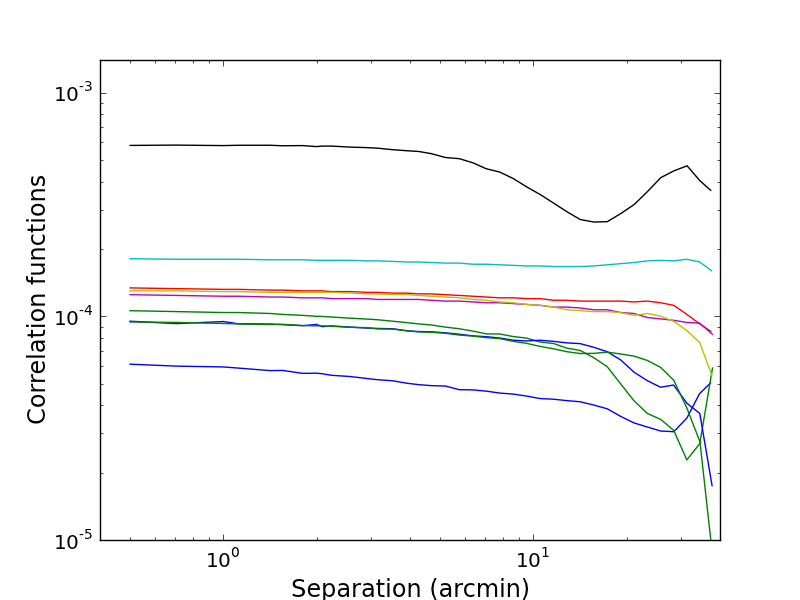}
\caption{\label{fig:atmos_xipm}For several simulated atmospheric PSF
  fields from the LSST PhoSim with 60s exposure times, we show the
  correlation function of PSF shapes defined as $\xi_+$ (Eq.~\ref{eq:xip}).  $\xi_-$ (Eq.~\ref{eq:xim}) is consistent with
zero and therefore not shown.  Different line colors are different
realizations with randomly-chosen atmosphere parameters.}
\end{center}
\end{figure}

We calculate the star ellipticities, $e$ (Eq.~\ref{eq:ellipticity}), and their correlation functions; 
examples 
are in Fig.~\ref{fig:atmos_xipm}.  The
zero-lag value of the correlation function is simply the variance of
the shapes, i.e., $\langle e_1^2 + e_2^2\rangle$ averaged over all
simulated stars in the exposure.  The salient features of this plot are
\begin{itemize}
\item There is substantial range in the amplitude and slope of this
  correlation function between individual realizations.
\item The amplitude of the ellipticity variance 
is relatively small, typically in the range
  $10^{-4}$ to $10^{-3}$.
\item The shapes are coherent to quite large scales, an effect that
  has been seen even for simulations of a larger area than shown here.
(See Fig.~\ref{fig:ps_cf} for an example of how the shape correlation
functions compare to those for lensing shear or optical PSFs.)
\end{itemize}

In practice, we adopted a functional form that can describe these 
atmospheric PSF correlations in a set of images
covering a larger $2\times 2$ deg$^2$ field, and on even larger scales in
PhoSim.  That functional form has two parameters: the
overall PSF ellipticity variance and a length scale
determining how quickly the correlations die off with scale.  For a
given exposure in the variable PSF experiment, two random numbers are 
chosen for those parameters with a flat distribution (also consistent with
the simulations) to determine a PSF ellipticity correlation
function.  The amplitude also scales inversely with the exposure time
and the telescope diameter.  For the GREAT3 simulations, we choose random
values of 
exposure
times of $60$--$180$s. 

The spatial variation of the 
PSF size also follows a similar correlation function. Typical fractional 
fluctuations in size are a few tenths of a percent.

\subsection{GalSim implementation of the atmospheric PSF}

To simulate the atmospheric PSF for the challenge, we use the GalSim 
software that takes a lensing shear power spectrum and uses it to simulate
galaxy shears\footnote{This is significantly less computationally intensive
than making a large number of PhoSim simulations directly, and in the
long-exposure limit we can make simple models that capture the
relevant physics at the level needed for this challenge.}.
The physics in the two cases is the same, except that
for the atmosphere, there is equal power in $E$ and $B$ modes, whereas
lensing only generates $E$ modes.  Moreover, the fluctuations in PSF
size across the FOV are sourced by the same physical source of the
$E$-mode anisotropies, so we can use the ``convergences'' from the
GalSim outputs as fractional changes in PSF size.  Using
this software requires us to convert the PSF anisotropy correlation
function that we use to a power spectrum, via numerical integration.  We tabulate the power spectra
for logarithmically-spaced values of correlation length; for some 
random value of correlation length, we use the nearest one for which
the power spectrum was tabulated.  The GalSim lensing code can then 
generate a random realization of a gridded shear field with
very large spatial extent (to avoid issues with cutoffs
in correlations at the edges of our image, \S\ref{subsec:sims:shear})
using the chosen power spectrum divided into half  
$E$-mode and half $B$-mode power.  We use this gridded shear field
along with simple assumptions described at the end of
\S\ref{subsubsec:sims:psf:atm} to generate the atmospheric PSF as a
function of position in the field of view.

\section{$B$-mode shape noise}\label{sec:shapenoise}

As described in \S\ref{subsec:sims:noise}, in order to maintain a
reasonable simulation volume for the GREAT3 challenge, we need a 
way to remove the intrinsic galaxy shape noise from the quantity of
interest, the reconstructed shear correlation function.  However, the
scheme described there of using 90-degree rotated galaxy pairs does 
not work for spatially varying shear fields (see, e.g.,
Appendix A of \citealt{2010arXiv1009.0779K}).  As in the GREAT10
challenge, we adopt a scheme to ensure 
that the intrinsic shape noise only shows up in the $B$ mode shear
correlation function, whereas the lensing shear is only $E$ mode.

However, the situation in GREAT3 is somewhat more complex because the
galaxy $p(|\varepsilon^{(s)}|)$ is determined for us by the galaxy training
sample that we are reproducing.  Whereas in GREAT10 it was possible to
generate a Gaussian random field of pure $B$-mode intrinsic
ellipticities with an appropriate variance, in GREAT3 we only have the
freedom to choose the orientations, not the ellipticity magnitudes, of
the galaxies which we simulate.  Without altering the
$p(|\varepsilon^{(s)}|)$ for our training sample, which we wish
to avoid doing as the $p(|\varepsilon^{(s)}|)$ is an important
characteristic of realistic galaxy populations for weak lensing, it is
impossible to avoid some leakage of shape noise into $E$-modes where
it increases the uncertainty on GREAT3 submission results.

The extent of this $B$-mode leakage can be reduced 
using a prescription we now describe.  First, an estimate of
$\varepsilon$ is made for every galaxy in the COSMOS training sample,
using the second moments of high resolution images of the
model fits described in Appendix~\ref{sec:galaxies}.
Taking these estimates of $\varepsilon^{(s)}$ for the training sample, we
calculate the variance in each component of ellipticity, ${\rm
  Var}[\varepsilon^{(s)}_1]$ and ${\rm Var}[\varepsilon^{(s)}_2]$.  These determine
the variance $\sigma_I$ of the pure $B$-mode, constant power spectrum,
Gaussian random field to use as a `target' for the intrinsic galaxy shapes
in the simulations, $\sigma_I = {\rm Var}[\varepsilon^{(s)}_1] + {\rm
  Var}[\varepsilon^{(s)}_2]$.

We label as $b$ the resultant ellipticities for a given realization of
this target $B$-mode Gaussian field (using the $\varepsilon$
convention for ellipticity, see \S\ref{subsec:physics:shear}).
We seek to put down source galaxies from the training set with
$\varepsilon^{(s)}$ as close to $b$ as needed to ensure negligible
$B$-mode leakage.  
The $p(|b|)$ is, by
definition, a Rayleigh distribution with $\sigma = \sigma_I$.
Comparison with histograms of the training sample
$|\varepsilon^{(s)}|$ showed reasonable, but not perfect, agreement
between the distributions.  This provided encouragement that it might
be possible to generate a field of galaxies with nearly-pure $B$-mode
intrinsic shapes by appropriate selection from the training sample,
followed by rotation
(we are free to rotate our source galaxies to align their ellipticities with $b$).
The procedure adopted was then as follows:
\begin{enumerate}
\item For each simulation field, a realization of pure $B$-mode
  ellipticities is generated as a Gaussian random field, yielding a
  target ellipticity $b_j$ at each
  of $j=1.,\ldots,N$ galaxy positions in the field.  We note that here
  the subscript $j$ does not denote shear component.
\item A sample of $N$ galaxy models are drawn from the full training
  sample, with replacement.  These models have estimated ellipticities $\varepsilon^{(s)}_k$.
\item The ranked ordering of $b_j$ by ascending $|b_j|$ is determined
  by sorting; the ranked ordering of $\varepsilon^{(s)}_k$ by
  ascending $|\varepsilon^{(s)}_k|$ is determined similarly.
\item At each galaxy position with target ellipticity $b_j$
  the source galaxy for which $|\varepsilon^{(s)}_k|$ took the
  same ordered rank as $|b_j|$ is selected, and assigned to this position.
\item This source galaxy is then rotated so that its ellipticity
  $\varepsilon^{(s)}_k$ is aligned with $b_j$.
\end{enumerate}
This procedure yielded samples of
source galaxies with intrinsic ellipticities that were acceptably close to being
a pure $B$-mode signal, while maintaining the real $p(|\varepsilon^{(s)}|)$
from the training set.  Simulations using the COSMOS training sample
demonstrated a leakage into the $E$-mode that was a factor of 7-8 smaller
in variance
than the expected shot noise $\sigma^2_n$ due to noisy pixels
(e.g., \S\ref{subsubsec:metrics:variable}), which is a 
tolerable contribution to the overall uncertainty.

\section{Simulating shear fields on finite grids}\label{sec:finitesim}

The Fourier space analogue of the shear correlation function is the
power spectrum $P(k)$, which describes the variance of a shear field
in Fourier modes as a function of the angular wavenumber
$k=|{\bf k}|$ on the sky.

An approximate simulation of a random shear field according to a
specified power spectrum is straightforward using the Discrete Fourier
Transform (DFT).  Inherent in the approach is that the underlying shear must
be approximated as a Gaussian random field, and values of the shear
are provided only at grid points of fixed spatial separation $\Delta
x$, which we label $\gamma_{ij}$.

The DFT of these shears, $\tilde{\gamma}_{ij}$
for discrete wavenumbers $k_{ij}$, can be generated as complex random
variables subject to the constraint that $\left\langle
  |\tilde{\gamma}_{ij}|^2 \right\rangle = \left( \Delta k \right)^2
P(k_{ij})$, where $\Delta k$ is the grid spacing in Fourier space
($\Delta k = 2 \pi / L$, where $L$ is the spatial extent of the 
grid in real space, and we assume a square grid
for simplicity). Drawing Gaussian random
deviates so that
\begin{equation}
\tilde{\gamma}_{ij} = \Delta k \sqrt{\frac{P(k_{ij})}{2}}
\left[ N(0, 1) + {\rm i} N(0, 1) \right]
\end{equation}
satisfies this constraint.  Applying the inverse DFT to such a
realization yields $\gamma_{ij}$ with a power spectrum that can be
directly related to a periodic sample of the desired $P(k)$.  Provided
$\Delta x$ and $\Delta k$ are sufficiently small, this will be a good
approximation to the desired shear field.  For more details, see \\
\url{https://github.com/GalSim-developers/GalSim/blob/master/devel/modules/lensing_engine.pdf}.

\end{document}